\documentclass[%
 amsmath,amssymb,
 aps,
]{revtex4-2}

\usepackage{graphicx}
\usepackage{dcolumn}
\usepackage{bm}

\usepackage{tikz}
\usetikzlibrary{arrows,backgrounds}

\usepackage{float}

\usepackage{csquotes}
\usepackage{multirow}
\usepackage{tabularray}
\DeclareMathOperator*{\argmin}{arg\,min} 

\usepackage{comment}

\begin{document}

\preprint{APS/123-QED}

\title{Turbulence model augmented physics informed neural networks for mean flow reconstruction}

\author{Yusuf Patel}
 \email{yusuf.patel15@imperial.ac.uk}
\affiliation{%
 Imperial College London, Department of Aeronautics, London SW7 2AZ, UK\\
}%
\author{Vincent Mons}%
 \email{vincent.mons@onera.fr}
\author{Olivier Marquet}%
 \email{olivier.marquet@onera.fr}
\affiliation{%
 ONERA, The French Aerospace Lab, 92190 Meudon, France\\
}%
\author{Georgios Rigas}%
 \email{g.rigas@imperial.ac.uk}
\affiliation{%
 Imperial College London, Department of Aeronautics , London SW7 2AZ, UK\\
}%




\date{\today}

\begin{abstract}
Experimental measurements and numerical simulations of turbulent flows are characterised by a trade-off between accuracy and resolution. In this study, we combine accurate sparse pointwise mean velocity measurements with the Reynolds-Averaged Navier-Stokes (RANS) equations using data assimilation methods. Importantly, we bridge the gap between data assimilation (DA) using Physics-Informed Neural Networks (PINNs) and variational methods based on classical spatial discretisation of the flow equations, by comparing both approaches on the same turbulent flow case. Firstly, by constraining the PINN with sparse data and the under-determined RANS equations without closure, we show that the mean flow is reconstructed to a higher accuracy than a RANS solver using the Spalart-Allmaras (SA) turbulence model. Secondly, we propose the SA turbulence model augmented PINN (PINN-DA-SA), which outperforms the former approach by up to 73\% reduction in mean velocity reconstruction error with coarse measurements. The additional SA physics constraints improve flow reconstructions in regions with high velocity and pressure gradients and separation. Thirdly, we compare the PINN-DA-SA approach to a variational data assimilation using the same sparse velocity measurements and physics constraints. The PINN-DA-SA achieves lower reconstruction error across a range of data resolutions. This is attributed to discretisation errors in the variational methodology that are avoided by PINNs. We demonstrate the method using high fidelity measurements from direct numerical simulation of the turbulent periodic hill at $Re=5600$. 

\end{abstract}

\maketitle


%


\section{\label{sec:introduction}Introduction} 
Whilst the Navier-Stokes equations describe accurately the evolution of fluid flow in space and time, their direct numerical simulation (DNS) is intractable in turbulent regimes. As a compromise, industrial simulation is dominated by Reynolds-Averaged Navier-Stokes simulations (RANS), which govern the time-averaged flow quantities instead of unsteady time-varying values. However, the RANS equations are not in closed form and solving for these first-order statistics (mean flow quantities) requires knowledge of the second-order velocity statistics, known as the Reynolds stresses. This closure problem is tackled by use of turbulence models, most commonly Boussinesq linear eddy viscosity models (LEVM), including the Spalart-Allmaras (SA) model~\citep{Spalart:1992}, $k-\omega$~\citep{Wilcox:1988} and $k-\varepsilon$~\citep{Launder:1974} models. Whilst solutions of the RANS equations are computationally tractable, they are also less accurate than scale-resolving simulations. Although the turbulence models fully close the system of equations for the mean flow quantities, they typically rely on empirical estimates and parameter tuning, lacking accuracy and generality, as quantified in a review by Xiao {\it et al.}~\citep{Xiao:2019}. On the other hand, experimental methods may provide valuable information about real-world flows, which are nevertheless generally limited in terms of spatio-temporal resolution and do not give access to a full flow description. Error from sources, such as imperfect test parts and sensor noise may also impact measurement reliability. Particle Image Velocimetry (PIV) is often used to measure flow velocity but its resolution may be hardware restricted and limited to specific planes of interest (2D planes in a full 3D field), whilst pressure measurements are difficult to obtain in the bulk flow without being intrusive.

In conjunction with the large growth in the use of data-based methodologies to tackle a wide range of fluid mechanics problems~\citep{Duraisamy:2019,Brunton:2020}, data assimilation methods have enabled the augmentation of low-fidelity numerical simulations with experimental data, to reconstruct full and accurate mean flow descriptions from the latter that obey the RANS equations. Such methods thus allow, among others, to extract information not present in the original dataset, interpolate missing data by super-resolving the flow fields (such as coarse PIV fields), and infer missing fields (such as pressure and Reynolds stresses). Different methods exist for tackling such a reconstruction problem: variational methods~\citep{Singh:2016,Franceschini:2020}, Bayesian inference~\citep{Edeling:2014,Wu:2016}, real-state observer~\citep{Saredi:2021,Marquet:2022} and, amongst various machine-learning-based approaches, Physics-Informed Neural Networks (PINNs), which have recently been used in conjunction with RANS~\citep{Luo:2020,Eivazi:2022,Saldern:2022,Pioch:2023}. As further detailed in the following, both variational approaches and PINNs formulate data assimilation as an inverse problem, and more specifically as an optimisation one, where one aims to minimise the discrepancies between data and the reconstructed mean flow. However, variational approaches and PINNs vastly differ in the solution method of such an optimisation problem, such that one might expect significant differences between their respective results.

As above-mentioned, the variational approach formulates data assimilation as an optimisation problem, where the discrepancies between the reconstructed flow and data are minimised. This minimisation is performed in conjunction with the strong imposition of the flow governing equations, here the RANS equations, through the adjoint method. Both the RANS equations and their adjoint counterpart are solved based on classical Computational Fluid Dynamics (CFD) discretisation methods, such as the finite-element method that is employed here. Foures {\it et al.}~\citep{Foures:2014} notably applied variational data assimilation method to super-resolve the mean flow for a laminar 2D circular cylinder. By providing sparse, synthetic experimental data (data that has been generated using high-fidelity computational methods but treated as experimentally generated) on a rectangular grid, the mean flow was successfully super-resolved. Notably, a Helmholtz decomposition is applied to the divergence of the Reynolds stress tensor, separating it into a potential component and a divergence-free component, the latter forming the control vector in the optimisation procedure. This work was expanded by Symon {\it et al.}~\citep{Symon:2017} using the same variational data assimilation method but applied to a turbulent mean flow around an aerofoil at $Re = 13500$ using experimental two-component planar PIV data. The work by Franceschini {\it et al.}~\citep{Franceschini:2020} extends this work further. Instead of the Helmholtz decomposition, the Reynolds stress is broken down into a eddy viscosity term and a non eddy viscosity forcing term. The eddy viscosity term is solved by including the SA model to the governing equations, whilst the variational data assimilation is used to find the non eddy viscosity forcing and also a corrective forcing within the SA equation itself. This is intended to reduce the corrections required in regions where the LEVM is insufficient.
Whilst the present study focuses on flow reconstruction, it may still be worth mentioning that the above-mentioned corrections that are identified by variational data assimilation may then form a dataset to build machine-learning-based turbulence models, as performed in the field-inversion machine-learning approach ~\citep{Singh:2018,Holland:2019,Volpiani:2021}. 

As an alternative to variational data assimilation, we here also consider the PINN approach first presented by Raissi {\it et al.}~\citep{Raissi:2019}. In conventional neural network approaches, models are optimised by minimising a cost function, which is purely driven by error to data. The PINN framework enforces weakly the governing equations, by adding the residual PDE errors to the cost function. This is similar to the aforementioned variational approaches, although PINNs only softly constrain the governing equations. Constraining the optimisation with fully or partially known governing equations (i.e. Navier Stokes or RANS), in addition to measurement data, ensures that solutions are physical, improving model performance, reduces the possible solution space during optimisation and the dependence on high volumes of data to train models. This approach has already shown great promise across a range of inverse flow problems (water wave problem in Jagtap {\it et al}.~\citep{Jagtap:2022:2} and high speed and supersonic flows in Mao {\it et al.}~\citep{Mao:2020} and Jagtap {\it et al.}~\citep{Jagtap:2022}) and can even be extended to convolutional neural networks, as demonstrated in Gao {\it et al.}~\citep{Gao:2021} and Kelshaw {\it et al.}~\citep{Kelshaw:2022}. Eivazi {\it et al.}~\citep{Eivazi:2022} and Hasanuzzman {\it et al.}~\citep{Hasanuzzaman:2023} used PINNs that are constrained by the RANS equations to reconstruct the flow field within a subdomain, by providing data at the domain boundaries for all flow fields (from experimental PIV data). Sliwinski and Rigas~\citep{Sliwinski:2022} approached the same problem as Foures {\it et al.}~\citep{Foures:2014}, albeit using the PINN framework instead. Firstly, the Helmholtz decomposition is applied to the unknown divergence of the Reynolds stress tensor (forcing vector), as used in Foures {\it et al.}~\citep{Foures:2014}, to reconstruct the flow from mean velocity data (first order statistics) alone. Secondly, by considering the full Reynolds stress tensor and providing both sparse first- and second-order velocity statistical data, the pressure field was also successfully extracted. 
A similar approach was taken by von Saldern {\it et al.}~\citep{Saldern:2022}, using PINNs in combination with velocity data, to reconstruct the mean flow of a swirling turbulent jet. Pioch {\it et al.}~\citep{Pioch:2023} also uses PINNs, constrained by the RANS equations with the addition of turbulence models, to reconstruct the mean flow over a backward facing step. Additional research by Molnar {\it et al.}~\citep{Molnar:2023} propose a new PINN-based process for background-oriented Schlieren imaging (BOS), to reconstruct the density field reconstruction from experimental data, with greater accuracy than traditional BOS image processing algorithms.

One avenue to increase the capability of PINNs is through the development of novel PINN architectures. The PINN framework has been expanded with Conservative PINNs (Jagtap {\it et al}.~\citep{Jagtap:2020}) and Extended PINNs (Jagtap and Karniadakis~\citep{Jagtap:2021}), with further development by Shukla {\it et al.}~\citep{Shukla:2021}. These methods divide the domain into subdomains, each with its own PINN, increasing the representative capacity of the framework, as discussed in \citep{Hu:2022}. Additionally, Mishra and Molinaro~\citep{Mishra:2022} and De Ryck{\it et al.}~\citep{De:2023} complete error analysis on the use of (extended) PINNs. Alternatively, one can apply advancements in other data-assimilation techniques to the existing PINN framework, in order to improve on previous studies, as performed here. This study thus aims to demonstrate that similar strategies as proposed in variational approaches may be developed in the PINN framework, to achieve the reconstruction of turbulent mean flows from sparse mean-velocity measurements, extending the laminar PINN framework of Sliwinski and Rigas~\citep{Sliwinski:2022}.

We here introduce a turbulence model-augmented PINN (RANS with SA) following an approach that is inspired by the above-mentioned variational data assimilation study by Franceschini {\it et al.}~\citep{Franceschini:2020}: the Reynolds-stress term in the RANS equations is separated into a modelled component and a corrective one. The modelled component involves an eddy viscosity that is assumed to verify the SA equation, whilst the corrective component forms one of the PINN's output, along with the mean velocity, pressure and eddy viscosity fields. Compared to previous studies investigating PINNs in the context of RANS~\citep{Luo:2020,Eivazi:2022,Saldern:2022,Pioch:2023}, the present approach exploits traditional turbulence modelling, adding physical constraints on the reconstructed mean flow and facilitating the handling of sparse data as demonstrated in the following, whilst at the same time not being restricted by modelling deficiencies (e.g. non-compliance to the Boussinesq hypothesis) through the introduction of the correction term in the RANS equations. Moreover, the similarities between the present PINN-based methodology and the variational framework of Franceschini {\it et al.}~\citep{Franceschini:2020} allow to perform here detailed comparisons between PINN-based and variational data assimilation, which, to the extent of the authors' knowledge, has never yet been performed in the context of RANS and mean flow reconstruction, while only very few such comparisons for unsteady flow reconstruction may be found in the literature~\citep{Du:2023}.

The following sections will be set out as follows. Section \ref{sec:methodology} will begin with presentation of the mean flow RANS equations and detail the use of the SA turbulence model, which will be used in this work. This is followed by a description of the data assimilation procedure and its implementation using PINNs or a variational approach. Section \ref{sec:numerics} introduces the test case and its numerical set up: a turbulent periodic hill at $Re=5600$. This is followed by a description of the PINN architecture and training procedure in addition to the numerical setup for the variational data assimilation. Section \ref{sec:results} will present the data assimilation results of two parts. The first segment presents results of the PINN mean flow reconstruction, without use of a turbulence model. This will demonstrate the basic PINN's capabilities and its limitations. The second part will present results, showing the use of turbulence model augmentation (SA) with PINNs. These results will be compared first to the baseline PINN (without turbulence model) and then compared with results from the equivalent variational-based method. Section \ref{sec:conclusion} contains concluding remarks.

\section{\label{sec:methodology}Flow Equations and Data Assimilation Methodology}
In this article, the inverse problem is solved, where a mean (time-averaged) flow field is reconstructed from a set of partial observations. The observations are $N_{m}$ sparse high-fidelity mean velocity measurements at points $\mathbf{x_{m}} = (x_{m},y_{m},z_{m})$, i.e. $U_{m,i}$,  where $U_i$ is the $i$-th component of the mean  velocity and the subscript $m$ denotes the $m$-th sparse measurement. The reconstructed flow field is sought by solving the constrained optimisation problem, such that error to measurements, $J$, is minimised, where 
\begin{equation}
    J = \sum_{m=1}^{N_m}\sum_{i=1}^{3} \left ( U_{m,i} - \hat{U}_{i}(\mathbf{x_{m}}) \right )^2,
\label{eq:J}
\end{equation}
while simultaneously the reconstructed field satisfies the under-determined mean flow governing equations.
Here, $\hat{U}_{i}(\mathbf{x_{m}})$ is the reconstructed mean velocity at $\mathbf{x_{m}}$. Whilst the equations in this section are presented for general flows, the results demonstrate use for a two dimensional flow.

Firstly, the governing mean flow equations without (baseline) and with (SA) turbulence model are presented in Section \ref{sec:methodology:RANS}. Secondly, in Section \ref{sec:methodology:data}, the details of two data assimilation approaches used to solve the optimisation are described - an adjoint-based variational approach and a PINN approach.

\subsection{\label{sec:methodology:RANS} Formulation of the Reynolds stress closure}
By applying the Reynolds decomposition, an unsteady flow field can be separated into a mean and fluctuating component. Through time-averaging of the Navier-Stokes equations, one can obtain the steady RANS equations. The data-driven techniques used in this study will be applied to mean flow data, which satisfy these RANS equations
\begin{subequations}
\begin{equation}
    \frac{\partial U_i}{\partial x_i} = 0,
    \label{eq:RANS:mass}
\end{equation}
\begin{equation}
    U_{j}\frac{\partial U_i}{\partial x_j} + \frac{1}{\rho}\frac{\partial P}{\partial x_i} - \frac{\partial (2\nu S_{ij})}{\partial x_j} + \frac{\partial \overline{u_i'u_j'}}{\partial x_j} = 0,
    \label{eq:RANS:mom}
\end{equation}
\label{eq:RANS}%
\end{subequations}
where $P$ is the mean pressure field and $\overline{{u_i}'{u_j}'}$ are the Reynolds stress tensor terms. $S_{ij}$ is the mean strain rate tensor equivalent to $\frac{1}{2}\left (\frac{\partial U_i}{\partial x_j} + \frac{\partial U_j}{\partial x_i}\right )$. For a three-dimensional flow (3D), this results in a system of 4 equations with 10 unknowns (assuming no spanwise mean flow, this can be reduced down to 3 equations and 6 unknowns). Given the dependence of the mean flow solution, $\left (U_{i},P\right )$, on the terms of the unknown Reynolds stress tensor, $\overline{u_i'u_j'}$, determination of these terms is critical to the performance and capability of data-assimilation methods for reconstructing the mean flow field. The remainder of Section \ref{sec:methodology:RANS} will present formulations of the mean flow equations, such that data can be used to infer the Reynolds stress (closure) terms.

\subsubsection{Baseline formulation of mean flow equations}
To solve the inverse problem, an additional constraint is placed on the closure term to improve its inference. Firstly, the divergence of the Reynolds stress tensor (as it appears in the RANS equations \eqref{eq:RANS:mom}) is considered as a forcing vector (Reynolds forcing vector). This reduces the closure term from 6 individual Reynolds stresses to 3 individual forcing terms (or 2 forcing terms from 3 stresses with no spanwise mean flow). Subsequently, a Helmholtz decomposition is applied, as in Foures {\it et al.}~\citep{Foures:2014} and Sliwinski and Rigas~\citep{Sliwinski:2022}, 
\begin{equation}
    f_i \equiv -\frac{\partial \overline{{u_i}'{u_j}'}}{\partial x_j} = \frac{1}{\rho}\frac{\partial \phi}{\partial x_i} + f_{s,i},
\label{eq:force:helm}
\end{equation}
decomposing the forcing into a scalar, potential part ($\phi$) and a divergence-free, solenoidal, vectorial component ($f_{s,i}$). This latter condition (divergence-free) provides an additional equation. Introducing decomposition \eqref{eq:force:helm} into the RANS equations  \eqref{eq:RANS}, gives 
\begin{subequations}
\begin{equation}
    \frac{\partial U_i}{\partial x_i} = 0,
    \label{eq:RANS:helm:mass}
\end{equation}
\begin{equation}
    U_{j}\frac{\partial U_i}{\partial x_j} + \frac{1}{\rho}\frac{\partial \left ( P - \phi\right )}{\partial x_i}- \frac{\partial (2\nu S_{ij})}{\partial x_j} - f_{s,i} = 0,
    \label{eq:RANS:helm:mom}
\end{equation}
\begin{equation}
    \frac{\partial f_{s,i}}{\partial x_i} = 0.
    \label{eq:RANS:helm:div}
\end{equation}
\label{eq:RANS:helm}
\end{subequations}
In addition to the new equation, the Reynolds forcing vector now consists of 4 terms (3 in two dimensions), in the form of a solenoidal forcing vector and a scalar potential field. In this system, there are now 5 equations and 8 unknowns (4 equations and 6 unknowns in 2D). By combining the pressure and potential terms into a single scalar field ($P - \phi$), another unknown can be removed. The resultant flow field, $\left (U_{i},P-\phi,f_{s,i}\right )$ is composed of the mean velocity components, combined pressure and potential forcing term and the solenoidal forcing components. Providing mean velocity data alone, at discrete locations, is sufficient to close the system at those points. However, the $P - \phi$ term  can not be separated into its constituent parts and the pressure field cannot be determined. 

\subsubsection{Turbulence-model augmentation of mean flow equations}
Whilst the above approach combines data with a formulation of the governing equations to infer the unclosed divergence of the Reynolds stress tensor, the closure problem still poses a challenge away from data points, since an infinite number of potentially physically unrealisable solutions satisfy the under-determined governing equations. The following methodology is proposed to augment the aforementioned approach, by adding additional physical constraints.

When performing numerical simulations of the RANS equations, turbulence modelling is utilised to approximate the Reynolds stresses and close the equations. Typically, the Boussinesq approximation is used, where the Reynolds stresses are assumed to be isotropic and depend on the eddy viscosity $\nu_{t}$. However, neglecting the anisotropic component, as well as other modelling assumptions made in the calculation of $\nu_{t}$, introduce errors  which propagate to the mean flow solution. Instead, as applied in Franceschini {\it et al.}~\citep{Franceschini:2020}, one can decompose the Reynolds forcing into a modelled, isotropic part and a corrective component which can include the anisotropic part, in addition to any isotropic component not captured by the modelled part, such that
\begin{equation}
     f_{i} = \frac{\partial (2\nu_{t} S_{ij})}{\partial x_j} + f_{c,i} = \frac{\partial (2\nu_{t} S_{ij})}{\partial x_j} + \underset{f_{c,i}}{\underbrace{\frac{1}{\rho}\frac{\partial \phi}{\partial x_i} + f_{s,i}}}.
\label{eq:eddy:decomp}
\end{equation}
The first term represents the modelled forcing component and the second term the corrective forcing component, such that the mean flow solution matches the data (high-fidelity measurements). In this work, the Helmholtz decomposition is also applied to $f_{c,i}$. By augmenting the system with a turbulence model, one can solve for $\nu_{t}$. This work will use the one-equation Spalart-Allmaras turbulence model, which provides a governing equation for a pseudo eddy-viscosity variable $\tilde{\nu}$. The actual eddy viscosity $\nu_{t}$ may then be obtained from $\tilde{\nu}$ through an algebraic expression. The SA equations are given in Appendix \ref{sec:app:SA:BSL}. The decomposition \eqref{eq:eddy:decomp} has a closed (modelled) part, $\nu_{t}$, and unclosed (corrective) part, $f_{c,i}$. Equation \eqref{eq:eddy:decomp} can then be substituted into Equation \eqref{eq:RANS:mom} to derive the final SA augmented RANS equations,
\begin{subequations}
\begin{equation}
    \frac{\partial U_i}{\partial x_i} = 0,
    \label{eq:RANS:SA:helm:mass}
\end{equation}
\begin{equation}
   U_{j}\frac{\partial U_i}{\partial x_j} + \frac{1}{\rho}\frac{\partial \left ( P - \phi\right )}{\partial x_i}- \frac{\partial (2\left (\nu + \nu_{t} \right ) S_{ij})}{\partial x_j} - f_{s,i} = 0,
    \label{eq:RANS:SA:helm:mom}
\end{equation}
\begin{equation}
    \frac{\partial f_{s,i}}{\partial x_i} = 0,
    \label{eq:RANS:SA:helm:div}
\end{equation}
\begin{equation}
    U_{j}\frac{\partial \tilde{\nu}}{\partial x_j} - S_{p} - S_{diff} - S_{c} - S_{d} = 0.
    \label{eq:RANS:SA:helm:SA}
\end{equation}
\label{eq:RANS:SA:helm}
\end{subequations}

Applying the turbulence model allows an isotropic component of Reynolds stresses to be approximated using the SA equation, leaving the anisotropic (and potentially an isotropic) component underdetermined. As a large portion of the Reynolds forcing is now modelled, data-assimilation techniques only need to apply smaller corrections, compared with Equation \eqref{eq:RANS:helm}, to close the solution. 

\subsection{\label{sec:methodology:data}Data assimilation techniques}
Using data assimilation techniques, to solve the constrained optimisation, a solution $\hat{U}_{i}(\mathbf{x}), \hat{P}(\mathbf{x})-\hat{\phi}(\mathbf{x}), \hat{f}_{s,i}(\mathbf{x}), \hat{\tilde{\nu}}(\mathbf{x})$ to the mean flow equations \eqref{eq:RANS:helm} or \eqref{eq:RANS:SA:helm} is sought, such that $J$, the measurement error \eqref{eq:J}, is minimised. In the specific case where turbulence model augmentation is not applied, as in Equation \eqref{eq:RANS:helm}, the eddy viscosity field $\tilde{\nu}(\mathbf{x})=0$. Two techniques are implemented to solve the constrained optimisation problem: a neural network approach, specifically the PINN approach from Sliwinski and Rigas~\citep{Sliwinski:2022} and an adjoint-based variational approach, used by Foures {\it et al.}~\citep{Foures:2014} and Franceschini {\it et al.}~\citep{Franceschini:2020}.

\subsubsection{\label{sec:methodology:data:PINN}Physics informed neural network}
\begin{figure}[t!]
    \centering
    \includegraphics[width=0.9\textwidth]{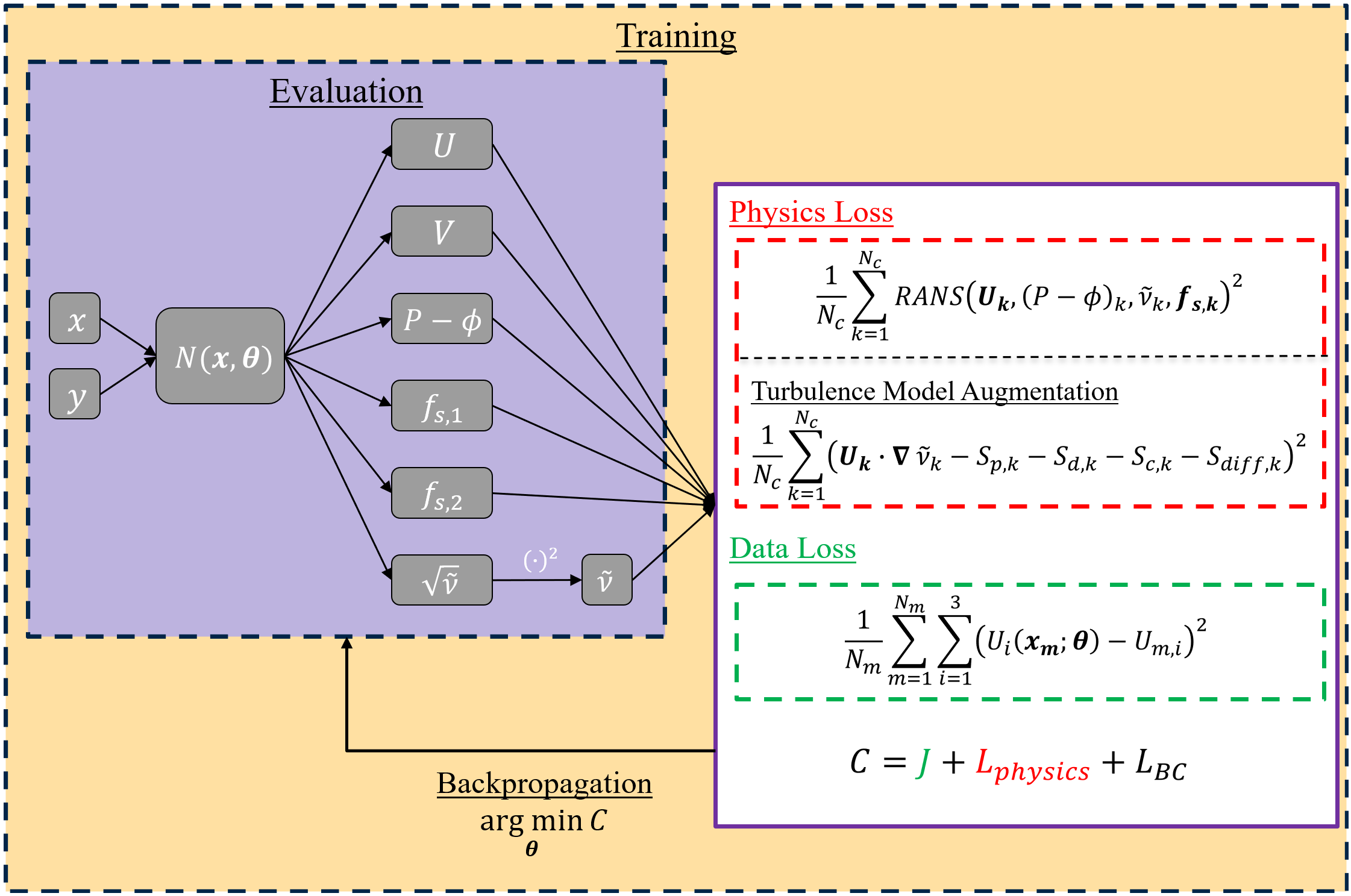}
    \caption{Structure of the Physics-Informed Neural Network. A continuous mapping between spatial coordinates and flow variables is approximated by a deep neural network, denoted by $N$, defined by a set of weights and biases, $\mathbf{\theta}$. High-fidelity coarse data measurements and the (turbulence model augmented) RANS equations constrain the training of the network. $k, m$ subscript the physics and data loss, respectively. $i$ represents the $i$-th component of a vector.}
    \label{fig:PINN}
\end{figure}

For a PINN, the solution $(\hat{U}_i,\hat{P}-\hat{\phi},\hat{f}_{s,i},\hat{\tilde{\nu}})(\mathbf{x;\theta})$, is defined by weights and biases, $\mathbf{\theta}$, which define a neural network. Figure \ref{fig:PINN} shows this PINN schematic. For any neural network, the optimal set of weights and biases, $\mathbf{\hat{\theta}}$, is sought by minimising some loss function $C$,
\begin{equation}
    \hat{\theta} = \argmin_{\mathbf{\theta}} C(U_i,P-\phi,f_{s,i},\tilde{\nu})(\mathbf{x;\theta}).
\label{eq:PINN:opt}
\end{equation}
In the case of PINNs, the loss function, $C$, is used to enforce the governing equations and the boundary conditions. Whilst these two terms are sufficient to define an optimisation for the forward problem, when using data assimilation (the inverse problem) the cost function must also include the measurement error term, $J$. To distinguish the use of PINNs for the inverse problem (as in this paper) compared to the forward problem, it will be referred to as PINN-DA hereon-in to emphasise the use of data. For the mean flow problem described in Section \ref{sec:methodology:RANS}, the loss function can be defined as
\begin{equation}
\begin{split}
    C =\frac{\lambda^{D}}{N_m}J 
    &+ \frac{\lambda^{P}}{N_c} \sum_{k=1}^{N_c} RANS \left ( (U_{i},P-\phi,f_{s,i},\tilde{\nu})(\mathbf{x_k;\theta}) \right )^{2}
    + \frac{\lambda^{P}}{N_c} \sum_{k=1}^{N_c} \left ( U_{j}\frac{\partial \tilde{\nu}}{\partial x_j} - S_{p} - S_{diff} - S_{c} - S_{d}\right )(\mathbf{x_k;\theta})^{2} \\
    &+ \frac{\lambda^{B}}{N_b} \sum_{b=1}^{N_b} BC \left ( (U_{i},P-\phi,f_{s,i},\tilde{\nu})(\mathbf{x_b;\theta}) \right )^{2}
    + \frac{\lambda^{R}}{2}\frac{1}{N_c} \sum_{k=1}^{N_c}\sum_{i=1}^{3}f_{s,i}(\mathbf{x_k;\theta})^2,
\end{split}
\label{eq:PINN}
\end{equation}
where the governing laws are evaluated at $N_c$ collocation points (subscripted by $k$), and boundary conditions are imposed at $N_b$ points along the boundary (subscripted by $b$). The data loss is subscripted by $m$. $RANS$ is the residual of the mean flow equations (\eqref{eq:RANS:helm} or \eqref{eq:RANS:SA:helm}) and $BC$ is the error to the imposed boundary conditions. To evaluate the governing laws, the gradients in the RANS equations are calculated accurately to machine-precision using automatic differentiation (leveraging the chain rule to trace the derivatives of all the constituent operations in the mapping $N(\mathbf{x,\theta})$ from Figure \ref{fig:PINN}). The data term, physics term and boundary condition terms are weighted by factors $\lambda^{D}, \lambda^{P}$ and $\lambda^{B}$, respectively. Selection of these weights is discussed in Section \ref{sec:res:conv}.

The decomposition of the forcing into a modelled and corrective component \eqref{eq:eddy:decomp} is not unique, as discussed in Foures {\it et al.}~\citep{Foures:2014}. As a result, additional regularisation is applied to ensure uniqueness of the distribution between modelled and corrective forcing during optimisation. For the PINN-DA, an $L_2$ regularisation is used for its effectiveness and ease of implementation. This regularisation term (also evaluated at collocation points), weighted by $\lambda^{R}$, controls the magnitude of the corrective forcing field and subsequently the decomposition between modelled and corrective forcing. By increasing the penalisation of forcing magnitude, the importance of the modelled forcing component will grow and the governing equations will tend towards the RANS-SA equations (no corrective forcing).
The effect of regularisation and importance of $\lambda^{R}$ is demonstrated in Appendix \ref{sec:app:reg:PINN}. In the case without turbulence-model augmentation (effectively $\tilde{\nu}(\mathbf{x})=0$), $\lambda^{R}=0$ and the SA equation can be removed from the loss function. The total loss $C$ function \eqref{eq:PINN:opt} is minimised using backpropagation by iteratively adjusting the weights and biases based on the gradient of loss with respect to the network parameters, ${\partial C} / {\partial \mathbf{\theta}}$. 

\subsubsection{\label{sec:methodology:data:var} Variational assimilation}
In the variational approach, the corrective forcing $f_{s,i}$ is adjusted such that the measurement error, $J$, is minimised, constrained by the RANS equations \eqref{eq:RANS:SA:helm}. The solution to this minimisation problem is given by the stationary points of the Lagrangian
\begin{equation}
\begin{split}
    \mathcal{L}\left ( \left [ U_{i}, P-\phi, \tilde{\nu}, f_{s,i} \right ], \left [ U_{i}^{*},P^{*},\tilde{\nu}^{*} \right ]  \right ) = J 
    &- \left < P^{*}, \frac{\partial U_i}{\partial x_i}\right > 
    - \left < U_{i}^{*}, U_{j}\frac{\partial U_i}{\partial x_j} + \frac{1}{\rho}\frac{\partial \left ( P - \phi\right )}{\partial x_i}- \frac{\partial (2\left (\nu + \nu_{t} \right )S_{ij})}{\partial x_j} - f_{s,i}\right > \\
    &- \left <\tilde{\nu}^{*}, U_{j}\frac{\partial \tilde{\nu}}{\partial x_j} - S_{p} - S_{diff} - S_{c} - S_{d}\right >.
\end{split}
\label{eq:var}
\end{equation}
The inner product is defined as
\begin{equation}
     \left <\mathbf{a}, \mathbf{b}\right > = \int_\Omega \mathbf{a}\cdot\mathbf{b} d\Omega,
\label{eq:innerproduct}
\end{equation}
where $\mathbf{a, b}$ are two spatial fields and $\Omega$ is the volume over which the field is defined. $U_{i}^{*},P^{*},\tilde{\nu}^{*}$ are the Lagrange multipliers (or adjoint variables) enforcing the constraints. By setting the variation of the Lagrangian with respect to the direct flow variables $U_i,P-\phi,\tilde{\nu}$, one can derive the adjoint equations
\begin{subequations}
\begin{equation}
    \frac{\partial U_{i}^{*}}{\partial x_i} = 0,
    \label{eq:adj:mass}
\end{equation}
\begin{equation}
   U_{j}^{*}\frac{\partial U_{j}}{\partial x_i} - U_{j}\frac{\partial U_{i}^{*}}{\partial x_j} - \frac{\partial P^{*}}{\partial x_i} - \frac{\partial (2\left (\nu + \nu_{t} \right ) S_{ij}^{*})}{\partial x_j} + \tilde{\nu}^{*}\frac{\partial \tilde{\nu}}{\partial x_i} + \frac{\partial }{\partial x_j}\left ( \tilde{\nu}^{*}\partial_{\nabla U_{i}}\left (S_p + S_d\right ) \right ) = \frac{\partial J}{\partial U_{i}},
    \label{eq:adj:mom}
\end{equation}
\begin{equation}
    -U_{j}\frac{\partial \tilde{\nu}^{*}}{\partial x_j} + 2\frac{\partial \nu_{t}}{\partial \tilde{\nu}}\frac{\partial U_{i}^{*}}{\partial x_j}S_{ij} - \tilde{\nu}^{*}\frac{\partial (S_{p} + S_{d})}{\partial \tilde{\nu}} + \frac{\partial }{\partial x_j}\left ( \tilde{\nu}^{*}\partial_{\nabla \tilde{\nu}}\left (S_c + S_{diff}\right ) \right ) = 0.
    \label{eq:adj:SA}
\end{equation}
\label{eq:adj}
\end{subequations}
Furthermore, by calculating the gradient of Lagrangian, $\mathcal{L}$, with respect to changes in forcing, $\frac{\partial \mathcal{L}}{\partial f_{s,i}}$, one finds $\frac{\partial \mathcal{L}}{\partial f_{s,i}} = \frac{dJ}{df_{s,i}} = U_{i}^{*}$. Specifically, this means the gradient $\frac{d J}{df_{s,i}}$ used to set the minimisation direction for optimisation is the solution to the adjoint equations \eqref{eq:adj}.

As with the PINN-DA-SA, regularisation is required to ensure uniqueness of the corrective forcing. Furthermore, the measurements are sparse, inducing a pointwise forcing of the adjoint momentum equations \eqref{eq:adj:mom} according to $\frac{\partial J}{\partial U_{i}}=2\sum_{m=1}^{N_m}\delta(\mathbf{x}-\mathbf{x_{m}}) (  U_{i}(\mathbf{x_{m}}) - U_{m,i})$, which may lead to discontinues in the adjoint field and in the gradient $\frac{dJ}{df_{s,i}}$, and thus ultimately in the reconstructed forcing and corresponding mean-flow solution. This can be circumvented through $H_1$-like regularisation of the gradient, which consists in getting a smoothed gradient $\frac{dJ^R}{d\mathbf{f_s}}$ from the original gradient $\frac{dJ}{d\mathbf{f_s}}$ through the inversion of the following system
\begin{equation}
    \begin{pmatrix}
    \frac{1}{1+\beta}(I-\beta\Delta)\circ & \nabla\circ\\ 
    -\nabla\cdot \circ & 0
    \end{pmatrix}\begin{pmatrix}
    \frac{dJ^R}{d\mathbf{f_s}}\\ 
    \pi
    \end{pmatrix} = \begin{pmatrix}
    \frac{dJ}{d\mathbf{f_s}}\\ 
    0
    \end{pmatrix},
\label{eq:var:reg:grad:mat}
\end{equation}
where $\beta=l_r^{2}$ and $l_r$ may be interpreted as a filter length. In the following, $l_r$ is chosen as the spacing between measurement locations. Eq. \eqref{eq:var:reg:grad:mat} also involves a pseudo-pressure field $\pi$ which is introduced to preserve the divergence-free character of the gradient through the regularisation procedure. A complete derivation and details of the variational approach can be found in ~\citep{Foures:2014}, ~\citep{Franceschini:2020} and ~\citep{Mons:2022}.

It has to be noted that for high Reynolds numbers turbulent cases, the variational-based data assimilation requires the use of the SA turbulence model augmentation in order to numerically solve the discretised equations, unlike the PINNs which work (albeit with different accuracy) with and without the SA model. Previously, the laminar studies in  ~\citep{Foures:2014} have not required this augmentation. To distinguish this approach it shall be referred to as variational-DA-SA.

\subsubsection{Comparison: PINNs vs variational assimilation}
Equations \eqref{eq:PINN} and \eqref{eq:var} highlight the similarities as well as key differences between the variational-DA-SA and PINN-DA approaches. Firstly, whilst the variational-DA-SA directly adjusts the corrective forcing, through the gradient ${dJ} / {df_{s,i}}$, the PINN-DA indirectly adjusts this forcing through adjustment of the network parameters, $\mathbf{\theta}$. Secondly, by the nature of PINNs, a continuous mapping of the flow solution is obtained, which can be queried at any spatial location. Additionally, the governing equations are evaluated and enforced at discrete locations within the domain (collocation points) and the spatial gradients in this mesh-free approach are determined through automatic differentiation. In the variational-DA-SA approach, solution of the direct and adjoint equations  introduces mesh dependency, where the conservation laws are enforced across the entire domain, whilst spatial gradients are calculated through discretisation (here Finite Element Method). As a result, discretisation errors are introduced, which are not present in PINNs. PINNs softly constrain the conservation laws in their continuous form whilst these are hard constraints in their discrete form for the variational-DA-SA. Lastly, the forcing regularisation method applied to the two approaches are different. The PINN-DA uses an $L_2$-based regularisation which was found to be sufficient to produce a smooth forcing solution. However, for the variational-DA-SA, an $L_2$ approach with pointwise data enforcement leads to unsmooth, pointwise forcing, thus the use of a $H_1$-based method, as discussed in Section \ref{sec:methodology:data:var}. Appendix \ref{sec:app:reg} contains a detailed breakdown on the effect of regularisation on PINNs and the effect of different regularisers for the variational-DA-SA approach.

\section{\label{sec:numerics}Numerical Simulations and Model Setup}
The PINN and variational approaches described above are applied to a canonical turbulent test case, the flow over a periodic hill. This section will describe the details of the flow case and the numerical setup of the two methods employed.

\subsection{\label{sec:numerics:PH}The periodic hill set-up}
\begin{figure}[t]
  \centering
  \begin{tabular}[b]{lclclc}
    \small (a) & & \small (b) & & \small (c) & \\
    & \includegraphics[height=105pt,trim={0.5cm 0.6cm 4.25cm 0.35cm},clip]{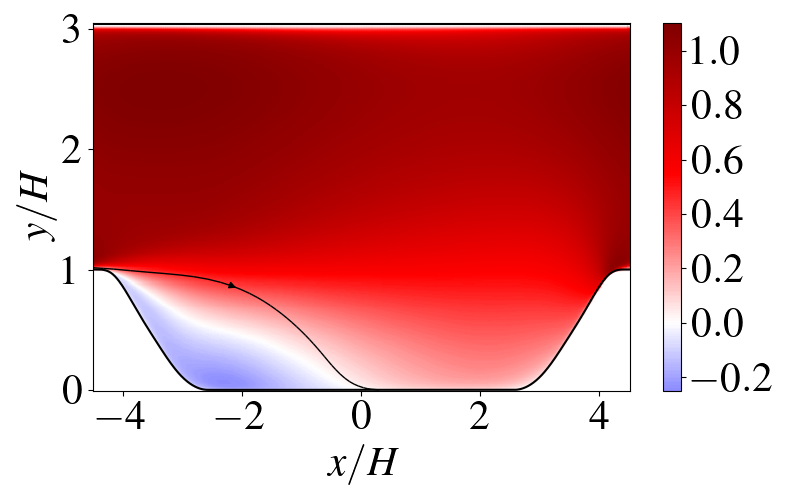} & & \includegraphics[height=105pt,trim={2.2cm 0.6cm 0.75cm 0.35cm},clip]{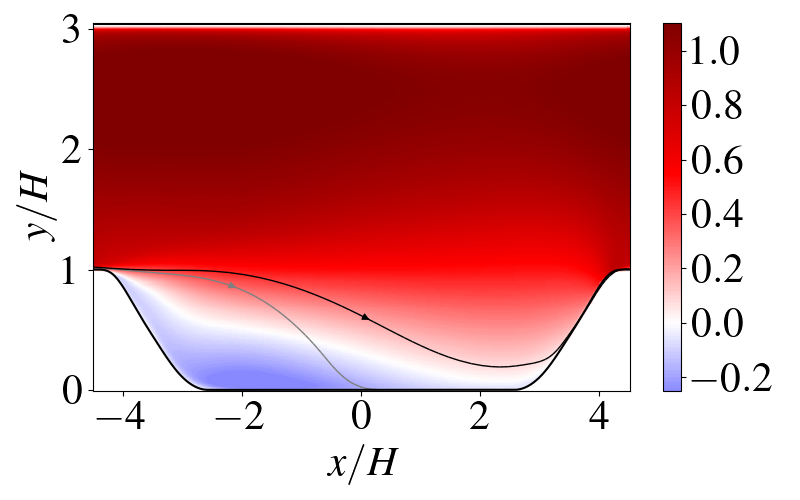} & & \includegraphics[height=105pt,trim={2.2cm 0.6cm 0.7cm 0.35cm},clip]{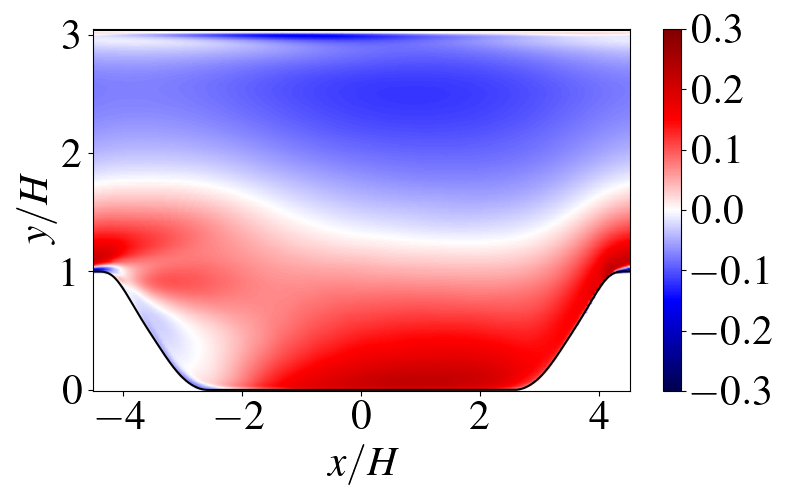}
  \end{tabular} 
  \caption{High-fidelity (DNS) solution of the flow over a periodic hill compared with the RANS solution using an SA model. The dividing streamline between the recirculation region and rest of the flow is shown. (a) Streamwise mean velocity - DNS; (b) streamwise mean velocity - RANS; (c) absolute error of streamwise velocity relative to DNS solution ($U_{err} = U_{DNS} - U_{RANS}$). For comparison, the DNS dividing streamline is shown in grey.}
  \label{fig:PH:DNS}
\end{figure}
The high-fidelity data for the periodic hill is obtained from the DNS database found in \footnote{The data is available from https://github.com/xiaoh/para-database-for-PIML} using the Incompact3D solver~\citep{Laizet:2011} and the simulation setup is detailed in Xiao {\it et al.}~\citep{Xiao:2020}. The DNS mean flow (time-averaged) solution, for streamwise velocity $U$ is shown in Figure \ref{fig:PH:DNS}(a). The streamline (solution to the streamfunction) dividing the recirculation region from the rest of the flow is also shown. The periodic hill is simulated at $Re=U_{b}H/{\nu}=5600$, where $H$ is the hill height, $U_b$ is the bulk velocity over the hill apex and $\nu$ the kinematic viscosity. Out of the several geometry configurations in the database, this work uses the canonical configuration with the domain length given by $L_{x}/{H} = 3.858\alpha + 5.142$ for which the hill stretch factor is $\alpha = 1.0$. At this Reynolds number, separation, recirculation and reattachment are present. The no-slip condition is enforced along the upper and lower walls, whilst periodicity is enforced in the streamwise direction. In order to maintain the required massflow and thus Reynolds number, body forcing is applied in the periodic streamwise direction. This is determined at each iteration, such that the Reynolds number matches the preset Reynolds number.

To analyse the effect of data resolution on mean flow reconstruction accuracy, the high-fidelity DNS data is spatially sampled on a square grid at nine data resolutions - $0.05 \leq \Delta L \leq 1.0$, where $\Delta L = \Delta x/H = \Delta y/H$. All results and figures hereon-in will compare the results from the $\Delta L = 0.5$ resolution, unless otherwise stated. The measurement locations are shown in Figure \ref{fig:colloc}. The choice of a sparse square sampling grid represents a velocity field as measured using methods such as Laser Doppler Velocimetry or PIV.

The low-fidelity RANS solution with the SA turbulence model is shown in Figure \ref{fig:PH:DNS}(b). Whilst the DNS reattachment occurs at $x/H=0$, the RANS solver massively overpredicts the size of the recirculation region, as shown. The error field likewise, in Figure \ref{fig:PH:DNS}(c), shows high error across the entire domain.

\subsection{PINN set-up\label{sec:numerics:PINN}}
The PINNs are implemented using the DeepXDE Python library by Lu {\it et al.}~\citep{Lu:2021}, as in ~\citep{Sliwinski:2022}. This toolbox simplifies the process of building PINNs (with the Raissi {\it et al.}~\citep{Raissi:2019} paradigm). The code and data to run the PINN cases from this paper are available at \url{https://github.com/RigasLab/PINN_SA}.

All PINNs use the same architecture. The neural network was constructed as a multilayer perceptron with 7 fully connected hidden layers, each with 50 nodes.
The input layer consists of two nodes, representing spatial coordinates, $x/H$ and $y/H$. The output layer consists of varying number of nodes depending on the test case. The PINN initially contains five output nodes with two mean velocity components, the combined mean pressure and potential forcing field, $P-\phi$, and two solenoidal forcing nodes. When the SA turbulence model augmentation was used, an additional output node is required for $\tilde{\nu}$. Each node used a $\tanh$ activation function for its smoothness and its second order differentiability. This architecture was deemed to perform well across a range of data resolutions.
A grid search method was employed to find optimal hyper-parameters (number of nodes and layers, activation function and learning rate). The effect of hyper-parameters on reconstruction accuracy are included in Appendix \ref{sec:app:hyper}. For PINNs, the weights and biases defining the network are randomly initialised using the Glorot uniform algorithm. This contrasts the variational-DA-SA approach, in which an \enquote{initial guess} (the RANS-SA solution) was used, around which the system is initially linearised. 

\begin{figure}[t]
  \centering\includegraphics[height=160pt,trim={2.3cm 2.0cm 0.6cm 0.45cm},clip]{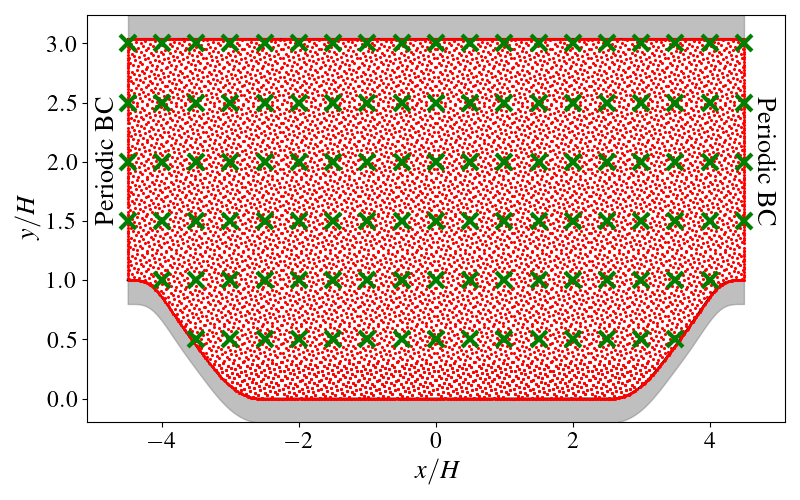}
  \caption{Turbulent periodic hill domain used for data assimilation, with $\Delta L=0.5$ data resolution (green crosses) used to compare the variational and PINN approaches. The collocation points, where the PINN evaluates residual error, are in red and were sampled across the domain using a Hammersley distribution.}
  \label{fig:colloc}
\end{figure}

For PINN optimisation, collocation points must be specified, in order to define where the governing laws are evaluated. In this work, the PINNs use 10000 collocations points which are distributed using a Hammersely distribution, such that there are sufficient points across all regions of the domain. This number of collocation points for the PINNs was selected after the reconstruction performance showed little sensitivity to further increases of the number of points. An additional 1000 points were distributed to the walls and periodic boundaries to enforce the boundary conditions. The collocation points (and domain) are shown in Figure \ref{fig:colloc}.
As in ~\citep{Sliwinski:2022}, a two step optimisation was used for the PINNs. Initially an ADAM optimisation is applied, followed by an L-BFGS-B phase. L-BFGS-B is a quasi-Newton method and thus its convergence depends on the initial guess. By using an ADAM optimiser first, one can reach an acceptable loss level to provide a good initial guess for the L-BFGS-B optimiser. Details on PINN convergence are discussed further in Section \ref{sec:res:conv}. The PINNs were trained on one NVIDIA RTX 6000 GPU.

\subsection{Modification to PINNs for Spalart-Allmaras augmentation\label{sec:numerics:PINN:mod}}
Using PINNs in combination with the SA turbulence model introduces additional complexity, which require modifications to the neural network architecture and the problem formulation to achieve convergence. 

Firstly, the SA transport equation contains $\frac{1}{d^2}$ terms in the production and destruction terms, where $d$ is the true wall distance, resulting in singularities at the wall, where $d=0$. This is not a problem for the finite element method (RANS and variational) where the residual is evaluated at a cell centre, which is not on the wall. For the PINN paradigm used, however, the governing equations are evaluated exactly at the wall boundaries and so this leads to singularities and subsequently failure to convergence. Consequently, the equation was reformulated by multiplying the SA equation with $d^2$ to eliminate the singularity. This new formulation can be found in Appendix \ref{sec:app:SA:Mod}. Problems arising from singularities when used with PINNs have also been mentioned in~\citep{Molnar:2023}.

The second change is to eliminate the effect of clipping negative values of $\tilde{\nu}$. The negative SA equation formulation used in the variational approach has distinct equations for positive and negative $\tilde{\nu}$ (detailed in Appendix \ref{sec:app:SA:BSL}) to ensure the transport equation is continuous across both positive and negative $\tilde{\nu}$. The final eddy viscosity, $\nu_{t}$, is then clipped at negative values of $\tilde{\nu}$ and set to zero, resulting in only positive values of $\nu_{t}$. Convergence of the PINNs is improved when the complexity caused by the discontinuity from clipping $\nu_{t}$ is removed. Thus, to enforce positive $\tilde{\nu}$, the output of the neural network is defined as $\sqrt{\tilde{\nu}}$. A transformation is then applied to the output, squaring the value, to produce a positive $\tilde{\nu}$. This is shown in Figure \ref{fig:PINN}.

\subsection{Variational assimilation set-up}\label{sec:var_setup}
The solutions of the RANS-SA system \eqref{eq:RANS:SA:helm} and its adjoint \eqref{eq:adj} are obtained through a finite-element method (FEM) spatial discretisation implemented in FreeFEM++~\citep{Hecht:2012}. Contrary to the presentation in Section \ref{sec:methodology:data:var}, the adjoint model is obtained following a discrete adjoint approach, where the Jacobian matrix that is associated to the RANS-SA equations is transposed. This Jacobian matrix is also used to invert the nonlinear RANS-SA equations based on a Newton method. Piecewise-linear functions that are enriched by bubble functions are used for velocity and pseudo-turbulent viscosity variables, while piecewise-linear functions are used for pressure. The FEM is known to be unstable at high Reynolds numbers. Accordingly, both streamline-upwind Petrov-Galerkin (SUPG) \citep{Brooks:1982,Franceschini:2020} and grad-div \citep{Olshanskii2009} formulations are here employed to stabilise the method. A low-memory Broyden–Fletcher–Goldfarb–Shanno (L-BFGS) \citep{Nocedal1980_mc} algorithm is used to exploit the gradient $\frac{dJ^R}{d\mathbf{f_s}}$ from Eq. \eqref{eq:var:reg:grad:mat} in order to perform the minimisation of the cost function $J$ in Eq. \eqref{eq:J}. As mentioned above, the first-guess for the optimisation procedure is $\mathbf{f_s}=\boldsymbol{0}$, i.e. the baseline solution of the RANS-SA equations. The full optimisation procedure was carried out on 28 cores Intel Xeon Broadwell E5-2680v4, 2.4 GHz. The total number of degrees of freedom in the RANS and adjoint simulations is around $2\times 10^5$. Further details on the implementation of the variational-DA-SA can be found in \citep{Franceschini:2020,Mons:2022}.

\section{\label{sec:results}Results}
\begin{figure}[t]
    \centering
    \begin{tabular}{lclclc}
    \small (a) & & \small (b) & & \small (c) & \\
    & \includegraphics[height=110pt,trim={0.4cm 0.5cm 0.35cm 0.25cm},clip]{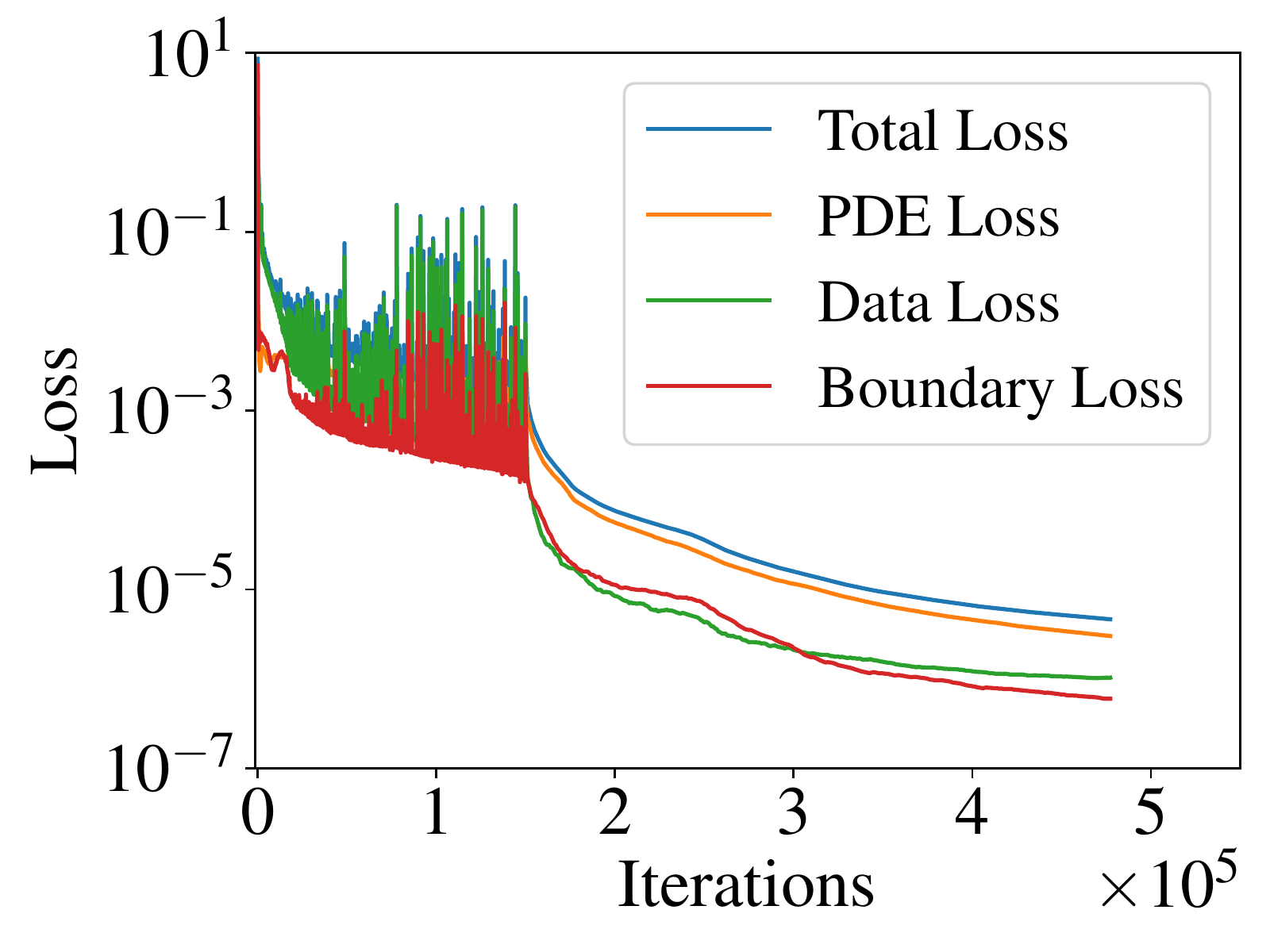} & & \includegraphics[height=110pt,trim={0.4cm 0.5cm 0.35cm 0.25cm},clip]{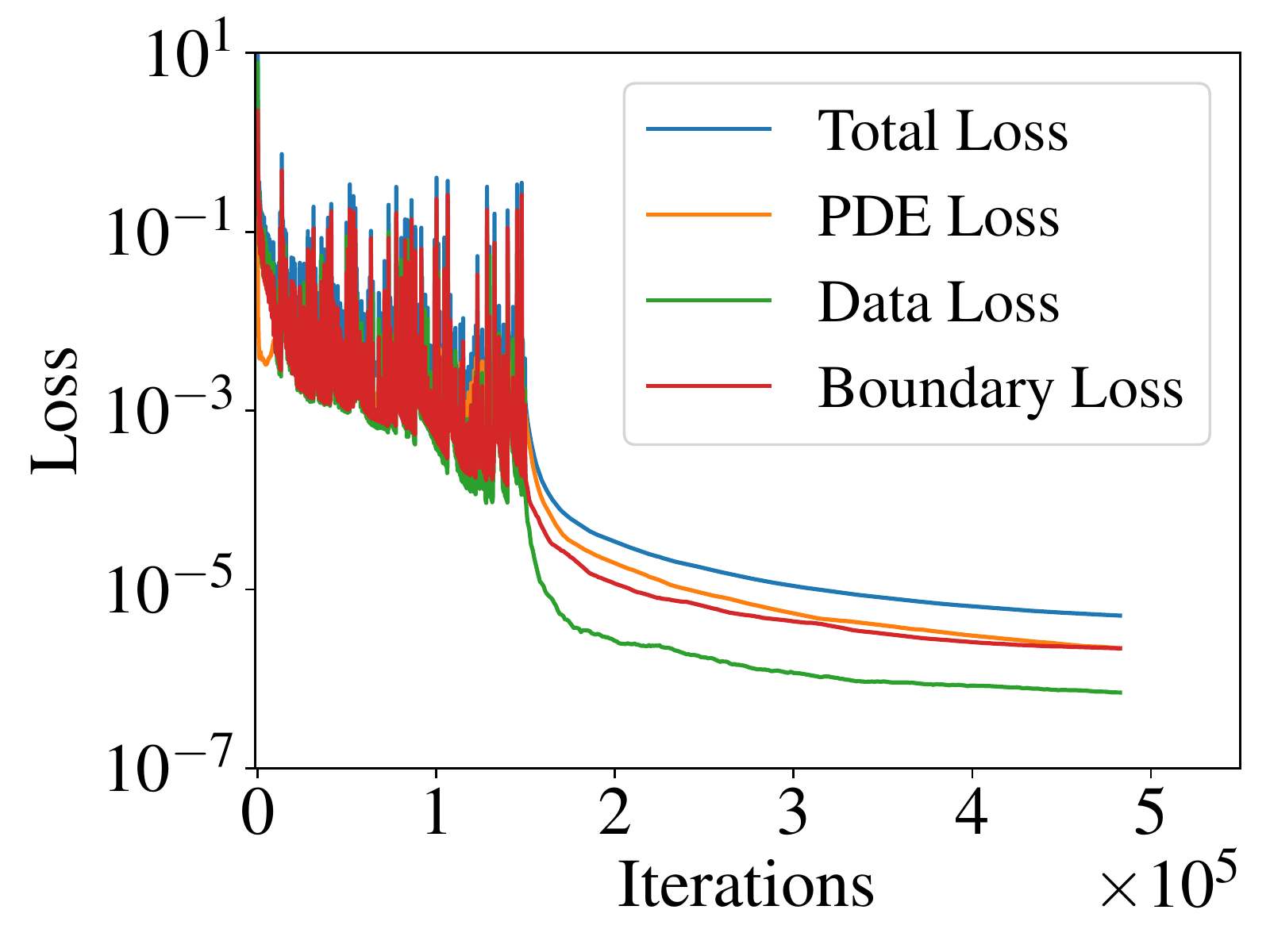} & & \includegraphics[height=110pt,trim={0.4cm 0.5cm 0.35cm 0.25cm},clip]{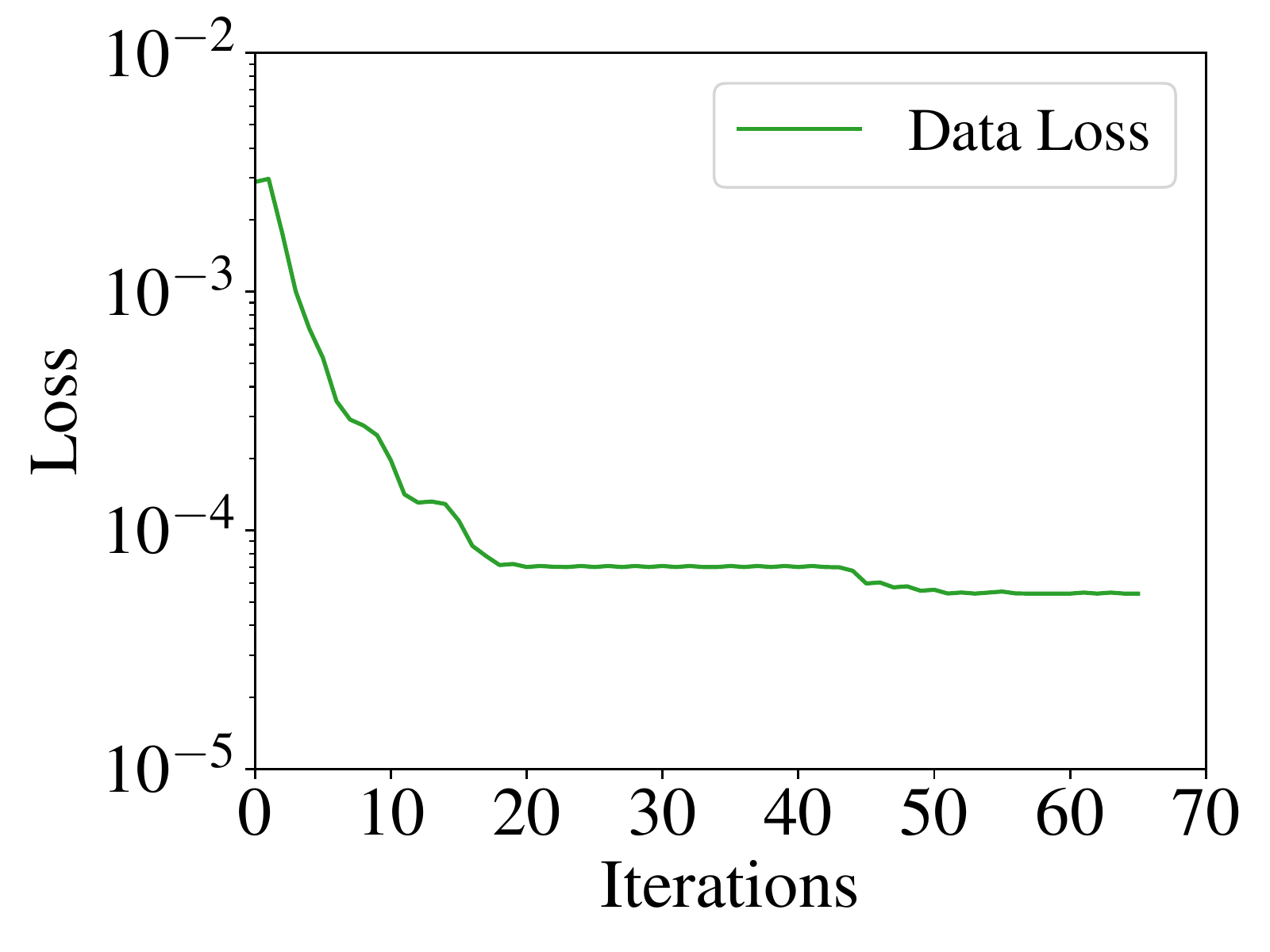}
    \end{tabular}
    \caption{Residual loss during optimisation for (a) PINN-DA-Baseline, (b) PINN-DA-SA and (c) Variational-DA-SA. For PINN optimisation (training), there are two distinct phases: the ADAM (stochastic gradient descent based) optimisation for 150000 iterations followed by the L-BFGS-B phase.}
    \label{fig:conv}
\end{figure}
\begin{table}[b]
\caption{Loss function weights used in PINN optimisation. For PINN-DA-Baseline, the $\tilde{\nu}$ and $\lambda^{R}$ weights can be ignored. Additionally, this regularisation weight is only valid for the $\Delta L = 0.5$ data resolution. For other resolutions, see Figure \ref{fig:eddy:regvsres}.}
\label{tab:weight}
\begin{ruledtabular}
\begin{tabular}{lccccccc}
\textrm{Weight}&
\textrm{$\lambda^{P}$}&
\textrm{$\lambda^{D}$}&
\textrm{$\lambda^{B} (U_{i})$ - Wall}&
\textrm{$\lambda^{B} (\tilde{\nu}, f_{s,i})$ - Wall}&
\textrm{$\lambda^{B}$ - Periodic}&
\textrm{$\lambda^{R}$}\\
\colrule
Value &
1 &
10 &
2.5 & 10 &
1 &
0.01\\
\end{tabular}
\end{ruledtabular}
\end{table}
We compare the mean flow reconstruction from sparse mean velocity measurements for the turbulent periodic hill case using three approaches: PINN-DA-Baseline (without turbulence model), PINN-DA-SA (with SA) and variational-DA-SA (with SA). Firstly, an evaluation of the PINN-DA-Baseline approach is presented. Secondly, we will show that significant reduction in reconstruction error is achieved by augmenting the PINN-DA-Baseline approach with the SA turbulence model (PINN-DA-SA). Thirdly, we compare it directly to the variational-DA-SA showing the increased accuracy of the PINN-DA-SA relative to the variational-DA-SA.

To evaluate the mean flow reconstruction accuracy, a volume weighted $L_{2}$ error will be used and is calculated as
\begin{equation}
    \varepsilon_{2} = \sqrt{\frac{1}{\Omega}\sum_{n=1}^{N}\delta\Omega_{n}\sum_{i=1}^{2} \left ( U_{n,i} - \hat{U}_{i}(\mathbf{x_{n}}) \right )^2},
    \label{eq:L2}
\end{equation}
where $\Omega$ is the volume of the domain, $\delta\Omega_{n}$ is the cell volume of each point used to determine the error (as defined by the DNS grid) and $N$ is the number of points across the domain. $U_{n,i}$ is the high fidelity, DNS mean velocity at $\mathbf{x_{n}}$ and $\hat{U}_{i}(\mathbf{x_{n}})$ is the corresponding reconstructed mean velocity. The variational solution is interpolated onto the DNS grid using a linear interpolation method. Additionally, the employed finite-element mesh corresponds to typical mesh sizes that are close to those in DNS.

\subsection{\label{sec:res:conv}Optimisation}
Figure \ref{fig:conv} shows the optimisation convergence of the PINN-DAs (a,b) and the variational-DA-SA approach (c). The PINN-DA loss convergence curve is decomposed into three loss components - PDE residual, data error and boundary condition error. For PINN-DA-SA the additional loss from the SA transport equation contributes also to the PDE loss. Each individual loss component appearing in Eq. \eqref{eq:PINN} was given initial weight values $\lambda$ based on the relative magnitudes of the flow variables. Thereafter, the weights were manually adjusted to find the combination with the lowest loss, as compiled in Table \ref{tab:weight}. The weight sensitivities are presented in Appendix \ref{sec:app:hyper}. During the first phase of the training, ADAM optimisation is performed for 150000 iterations. The second phase applies L-BFGS-B optimisation for 300000 iterations or until the model has sufficiently converged to the floating point tolerance. Figures \ref{fig:conv}(a,b) shows that during optimisation (for both PINN-DA methodologies), the ADAM phase reduces loss from the high value caused by the initial randomised state to approximately $10^{-3}$. During the ADAM optimisation phase, one can observe frequent jumps in the loss. This is an effect of the adaptive step size used in ADAM. The step size is inversely proportional to the exponential average of squared (loss) gradient (as described in ~\citep{Kochenderfer:2019}). As the loss flattens out, the adaptive step size becomes very large, causing the model to \enquote{jump} too far to a higher loss level. The L-BFGS-B phase is then used to fine tune the optimisation, reducing total loss to below $10^{-5}$. 

The convergence of the variational-DA-SA, seen in Figure \ref{fig:conv}(c) shows the data loss, corresponding to $J/N_{m}$, where $J$ is defined in \eqref{eq:J}. This data loss is equivalent to the corresponding curves in Figures \ref{fig:conv}(a,b). Whilst convergent, the data loss is an order of magnitude higher for the variational-DA-SA compared with the PINN-DAs. Whilst there is a vast difference in number of iterations between the PINN and variational approaches, an iteration is not equivalent across both methods. For PINNs, a single iteration involves evaluating the PINN at all collocation and data points, calculating the cost function and back-propagating the loss to update the model weights (and thus indirectly the flow solution). For the variational approach, in each iteration the direct and adjoint equations are solved after which the forcing solution is directly updated. An average PINN iteration takes approximately $0.84$ s/iteration, whilst for the variational approach this is approximately $55$ s/iteration. This difference in computational cost can be attributed to computational complexity, more mesh points in the variational method (than PINN collocation points), reduced number of degrees of freedom in the neural network (updating only weights and biases instead of the actual solution across the domain) and hardware.

\subsection{\label{sec:res:PINNB}PINN-DA-Baseline}
\begin{figure}[t]
  \centering
  \includegraphics[height=160pt,trim={0.5cm 0.6cm 0.25cm 0.4cm},clip]{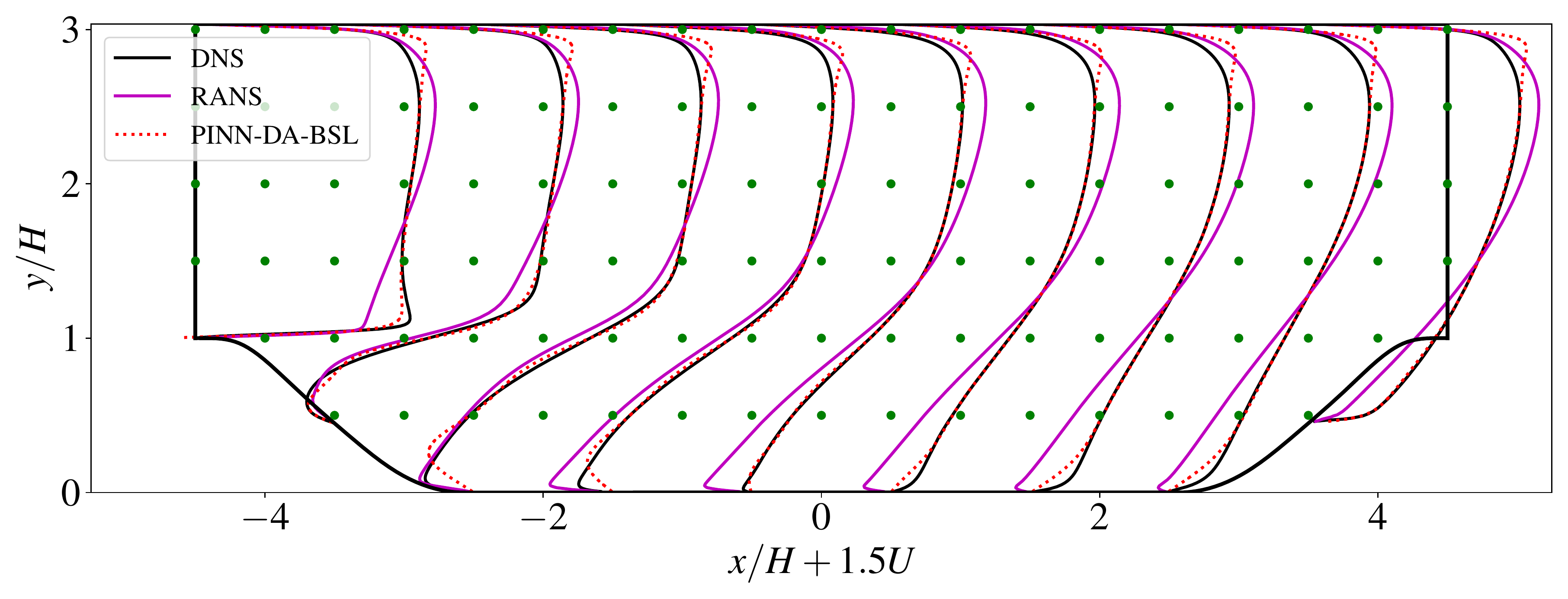}
  \caption{Streamwise velocity profiles for DNS, RANS-SA and PINN-DA-Baseline.}
  \label{fig:PINNB:profile}
\end{figure}
Here, the PINN-DA-Baseline approach is applied to reconstruct the mean flow velocity field when coarse mean flow velocity data is available with spacing $\Delta L=0.5$. This is compared against the DNS and RANS-SA velocity profiles in Figure \ref{fig:PINNB:profile}. The data assimilation of the high fidelity measurements has improved the prediction of the velocity field compared to RANS-SA. The total mean velocity $L_2$ error is $3.60\times 10^{-2}$ instead of the $8.99\times 10^{-2}$ error from RANS-SA. High accuracy of reconstruction is observed in the bulk flow domain. However, high mean velocity reconstruction error is observed in areas with high velocity gradients, such as near the walls and at separation. This is most apparent at the upper wall boundary layer. This near wall error accounts for over 40\% of the total $L_2$ error as the PINN-DA-Baseline fails to accurately capture high-gradient flow features. These trends are highlighted in the streamwise velocity contours in Figure \ref{fig:PH:comp}(a,d). At data points, the PINN-DA-Baseline reproduces the measurements accurately. Furthermore, in spite of the higher near-wall reconstruction errors, Figure \ref{fig:PH:comp}(a) shows improved reconstruction of the recirculation bubble, even with limited data in this area. This is most noticeable when comparing the PINN-DA-Baseline bubble from Figure \ref{fig:PH:comp}(a) with the RANS bubble in Figure \ref{fig:PH:DNS}(b). The provision of sparse high-fidelity datapoints in the PINN-DA-Baseline approach has enabled better prediction of the shape of the recirculation region. However, the error field makes the limitation of the PINN-DA-Baseline approach more apparent, with the near wall errors particularly noticeable.

The results of PINN-DA-Baseline show that the methodology used in ~\citep{Sliwinski:2022} for laminar flow can also be used for turbulent flows. In ~\citep{Sliwinski:2022}, the base formulation of the RANS equations (as in \eqref{eq:RANS}) was also tested. It was shown that by providing data for both first order (mean velocity) and second order (Reynolds stresses) statistics, the PINN-DA-Baseline was able to reconstruct the pressure field for the laminar case. This approach has been demonstrated in Appendix \ref{sec:app:rste}. With similar mean velocity reconstruction accuracy, this second approach was also able to infer the pressure field for this turbulent case.

\subsection{PINN-DA-SA\label{sec:res:PH:SA}}
This section contains results from the augmentation of the SA model to PINNs. First, a comparison between the PINN-DA-Baseline and PINN-DA-SA results will be presented, followed by a discussion on the  improved reconstruction when the SA model is used. Finally, the results of the variational data assimilation using SA (variational-DA-SA) are also presented, followed by a parametric analysis of the data resolution.

\subsubsection{PINN-DA-SA vs PINN-DA-Baseline\label{sec:res:PH:PINN:comp}}
\begin{figure}[t]
  \centering
  \begin{tabular}{clclclc}
    &            & \hspace*{0.5cm}PINN-DA-Baseline & &  PINN-DA-SA & & Variational-DA-SA\hspace*{1.0cm} \\
    & \small (a) & & \small (b) & & \small (c) & \\
    $\hat{U}$ & & \parbox[c]{142pt}{\includegraphics[height=91pt,trim={0.5cm 2.4cm 4.25cm 0.3cm},clip]{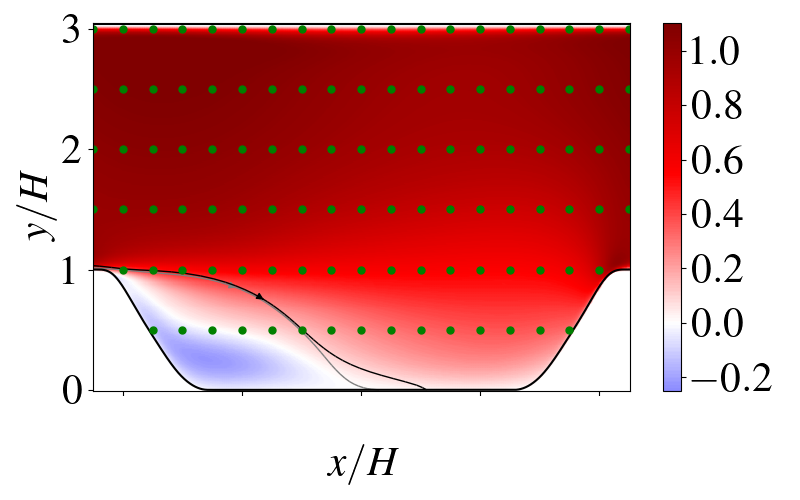}} & & \parbox[c]{126pt}{\includegraphics[height=91pt,trim={2.3cm 2.4cm 4.25cm 0.3cm},clip]{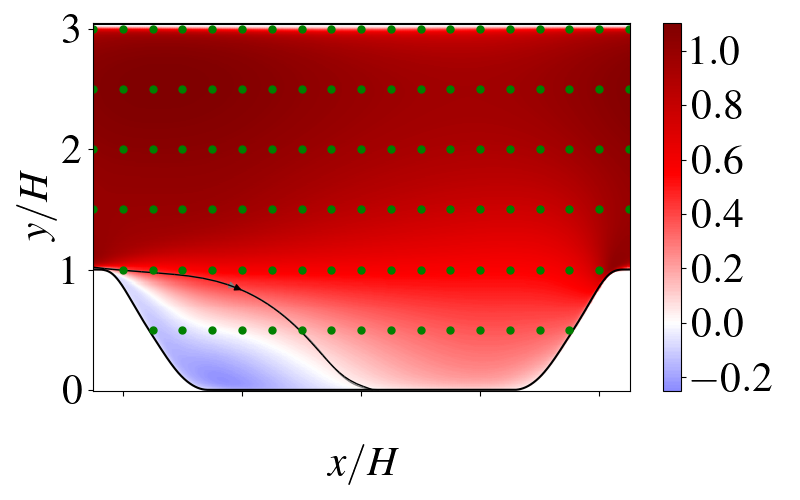}} & & \parbox[c]{158pt}{\includegraphics[height=91pt,trim={2.3cm 2.4cm 0.7cm 0.3cm},clip]{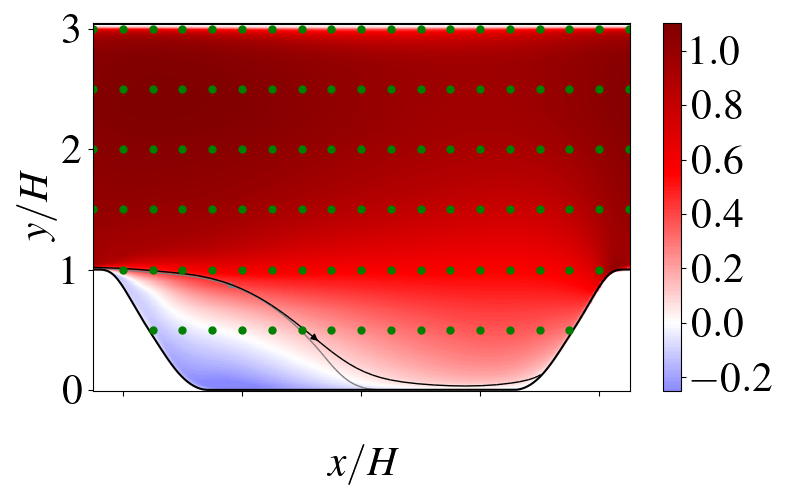}} \\
    & \small (d) & & \small (e) & & \small (f) & \\
    $U_{err}$ & & \parbox[c]{143pt}{\includegraphics[height=108pt,trim={0.5cm 0.6cm 4.25cm 0.3cm},clip]{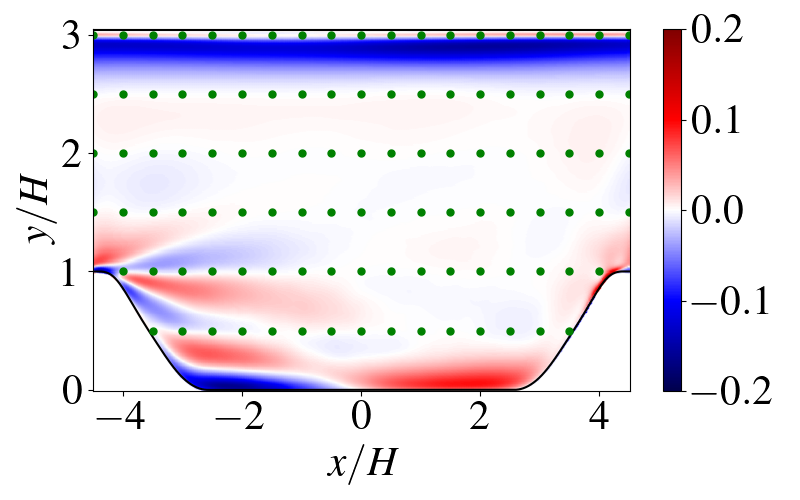}} & & \parbox[c]{126pt}{\includegraphics[height=108pt,trim={2.3cm 0.6cm 4.25cm 0.3cm},clip]{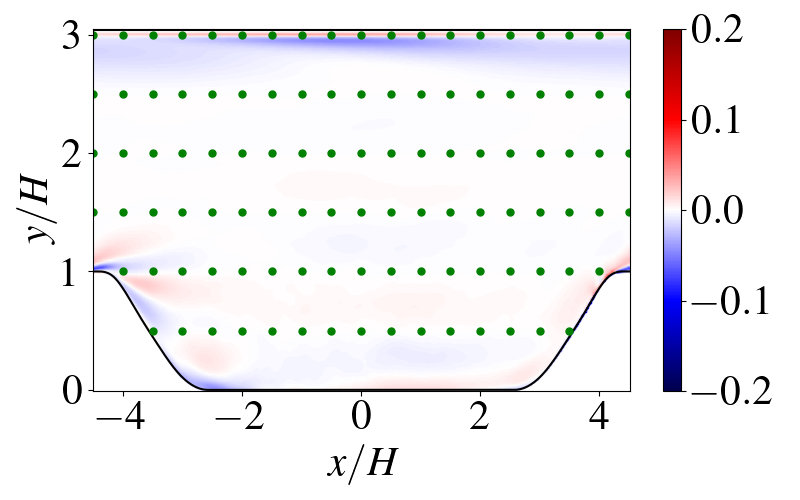}} & & \parbox[c]{159pt}{\includegraphics[height=108pt,trim={2.3cm 0.6cm 0.7cm 0.3cm},clip]{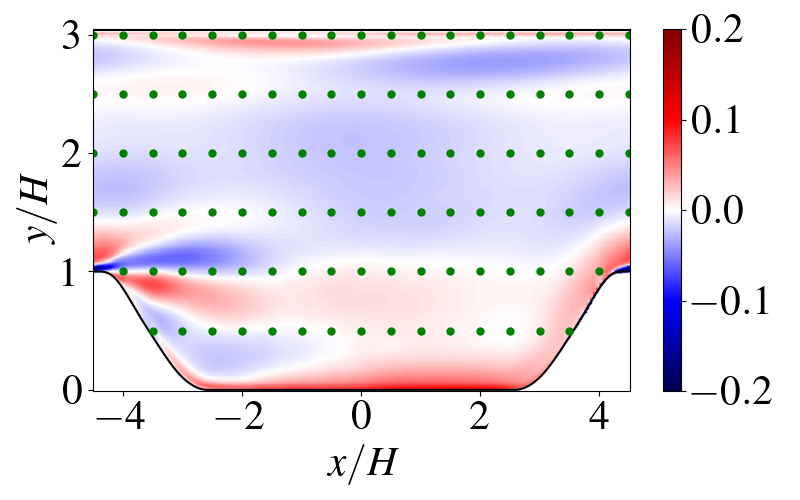}} \\
  \end{tabular}
  \caption{Reconstructed $U$ velocity component for the PINN-DA-Baseline (left), PINN-DA-SA (middle), variational-DA-SA (right) approaches. (a,b,c) - $U$ reconstruction. (d,e,f) - Absolute $U$ error. PINN-DA-SA performs best, significantly reducing the separation error seen in other approaches, whilst also reducing the near wall error seen in PINN-DA-Baseline. The dividing streamline (separating the recirculation region) is shown in (a,b,c) and compared with the DNS equivalent (in grey). Whilst the PINN-DA-SA recirculation region is indistinguishable, there is a small but noticeable difference for the PINN-DA-Baseline and variational-DA-SA solutions.}
  \label{fig:PH:comp}
\end{figure}
Figure \ref{fig:PH:comp}(a,d) shows the PINN-DA-Baseline and Figure \ref{fig:PH:comp}(b,e) the PINN-DA-SA reconstruction results. Streamwise velocity and error contours relative to the DNS solution are shown, with similar trends holding for the vertical velocity component (not shown here). Whilst the error field topologies are qualitatively similar, the  magnitude error is significantly reduced. For PINN-DA-SA, the recirculation bubble is almost indistinguishable from the DNS one. As a consequence of introducing the Spalart-Allmaras model, reconstruction error has reduced by 63\% ($\varepsilon_{2} = 1.35\times 10^{-2}$). For the bulk flow region above the hills and away from the walls (between $y/H=1.5$ and $y/H=2.5$) and at datapoints, the PINN-DA-SA and PINN-DA-Baseline show similarly low reconstruction errors. This is expected in this region given the lack of mean shear and production of turbulent kinetic energy. 

Comparing the PINN-DA-Baseline error against the PINN-DA-SA highlights the effect of augmenting the PINN with the SA turbulence model. The high mean flow error along the upper wall has largely reduced indicating that use of the SA model has enabled a better reconstruction of the near-wall region. There are also similarly significant improvements along the lower wall both at separation but also after reattachment. The improvements of PINN-DA-SA compared to PINN-DA-Baseline corroborate well with the original design purpose and performance of the SA turbulence model, which shows improved performance for turbulent boundary layers in adverse pressure gradients. 

\subsubsection{Why PINN-DA-SA?}
In order to understand the difference in performance between PINN-DA-Baseline and PINN-DA-SA, we examine the individual terms of the cost function, $C$, for each approach. This consists of the measurement error term (data loss) and the enforcement of the governing equations (PDE loss). 
 
Firstly, we examine the data loss component. Inspection of Figure \ref{fig:PH:comp}(d) reveals that the PINN-DA-Baseline reconstructs the mean velocity measurements accurately. This is comparable to the performance of the PINN-DA-SA, with the convergence plots in Figure \ref{fig:conv}(a,b) showing similar converged data loss values. Away from the data points, the reconstruction error increases substantially for the PINN-DA-Baseline. This is most apparent along the upper wall where a high error region exists above the data at $y/H = 2.5$. Given that the reconstruction error is low at the data points for PINN-DA-Baseline (and also similar to PINN-DA-SA), further decrease of the error to measurements, $J$, does not reduce the mean velocity error away from datapoints, where most of the reconstruction error lies. 

\begin{figure}[t]
  \centering
  \begin{tabular}[b]{lclc}
    & \hspace*{0.5cm}PINN-DA-Baseline & & PINN-DA-SA\hspace*{1.0cm} \\
    \small (a) & & \small(b) & \\
    & \includegraphics[height=140pt,trim={0.4cm 0.6cm 4.1cm 0.3cm},clip]{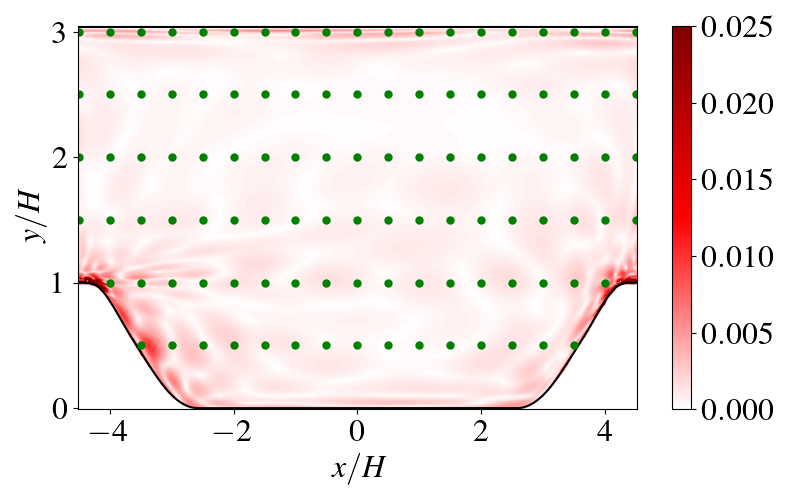} & & \includegraphics[height=140pt,trim={1.9cm 0.6cm 0.6cm 0.3cm},clip]{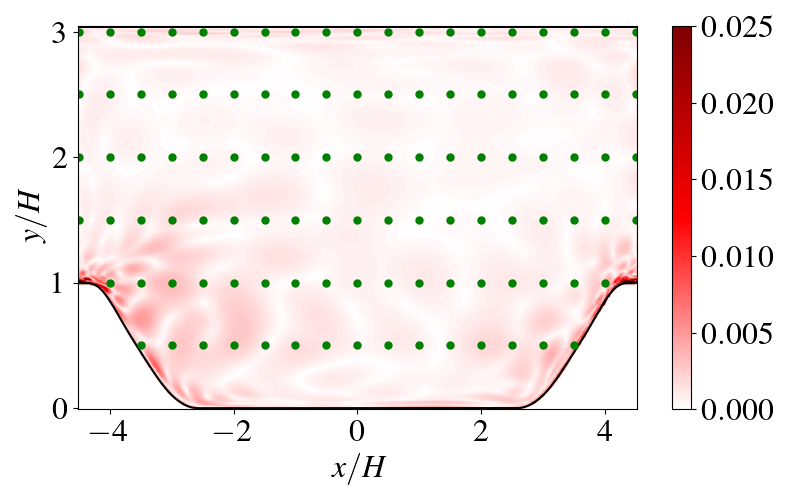} \\
  \end{tabular}
  \caption{PDE residual error (x-momentum and y-momentum). (a) PINN-DA-Baseline and (b) PINN-DA-SA. The residual error between approaches are comparable despite the considerable improvements in mean velocity reconstruction error demonstrated by the PINN-DA-SA.}
  \label{fig:PH:res}
\end{figure}

Secondly, one can consider the other component of the cost function, the residual PDE error. Figure \ref{fig:PH:res}(a) shows the PINN-DA-Baseline momentum residual error (for x- and y-momentum equations). The PDE residual error contour shows that there is a limited correlation between mean velocity reconstruction error and residual error from the conservation laws - most apparent along the upper and lower walls. A similar analysis for the PINN-DA-SA corroborates this trend (Figure \ref{fig:PH:res}(b)). The total residual PDE loss from both PINN-DA-Baseline and PINN-DA-SA (seen in Figure \ref{fig:conv}) are of similar magnitudes (approx. $10^{-5}$) and the residual PDE fields are comparable, despite the significant differences in reconstruction accuracy between the two PINN-DA approaches. This indicates that further reduction in the residual PDE error does not lead to increase in reduction in reconstruction error and thus cannot be attributed as the cause of the difference between PINN-DA-Baseline and PINN-DA-SA.

The difference between PINN-DA-Baseline and PINN-DA-SA can be consequently attributed to the effect of the turbulence model augmentation. Due to the unclosed (underdetermined) nature of the RANS equations, there are infinite solutions to the forcing (divergence of Reynolds stress tensor) in the mean flow equations. Providing high-fidelity data measurements \enquote{anchors} the PINN-DA solution at these discrete data points. However, away from them, there are still many candidate solutions to the mean flow, as the governing equations are still under-determined. These solutions all satisfy the governing equations and thus all have low residual PDE error. There are consequently no terms within the optimised cost function (Eq. \eqref{eq:PINN}) that can be minimised further, to reduce the mean velocity reconstruction error across the domain for PINN-DA-Baseline. The current optimisation for PINN-DA-Baseline is converged towards a solution that obeys the under-determined RANS equations. This is also briefly mentioned in Foures {\it et al.}~\citep{Foures:2014}. The SA model adds physics constraints that limit further the infinite number of admissible solutions far from the measurement points and as a direct consequence lead to improved reconstruction.

\subsubsection{Variational-DA-SA results and comparison with PINNs}
The PINN-DA-SA can be directly compared to the variational-DA-SA, which also uses SA augmentation, since both methods use the same data and physics constraints. Figures \ref{fig:PH:comp}(c,f) show the reconstructed mean streamwise velocity and the respective absolute error for the variational-DA-SA. As with all the data assimilation results presented, the key flow features (separation, recirculation bubble and reattachment) are well reconstructed. As with both PINN-DA methodologies, the variational-DA-SA reconstructs the shape of the recirculation bubble more accurately than the RANS-SA, despite the limited measurements in this region. The reconstruction error matches the features seen in the PINN-DA results as well as in \citep{Foures:2014}, with high error between data points, near walls and at the initial separation.

\begin{figure}[t]
  \centering
  \includegraphics[height=160pt,trim={0.5cm 0.6cm 0.25cm 0.4cm},clip]{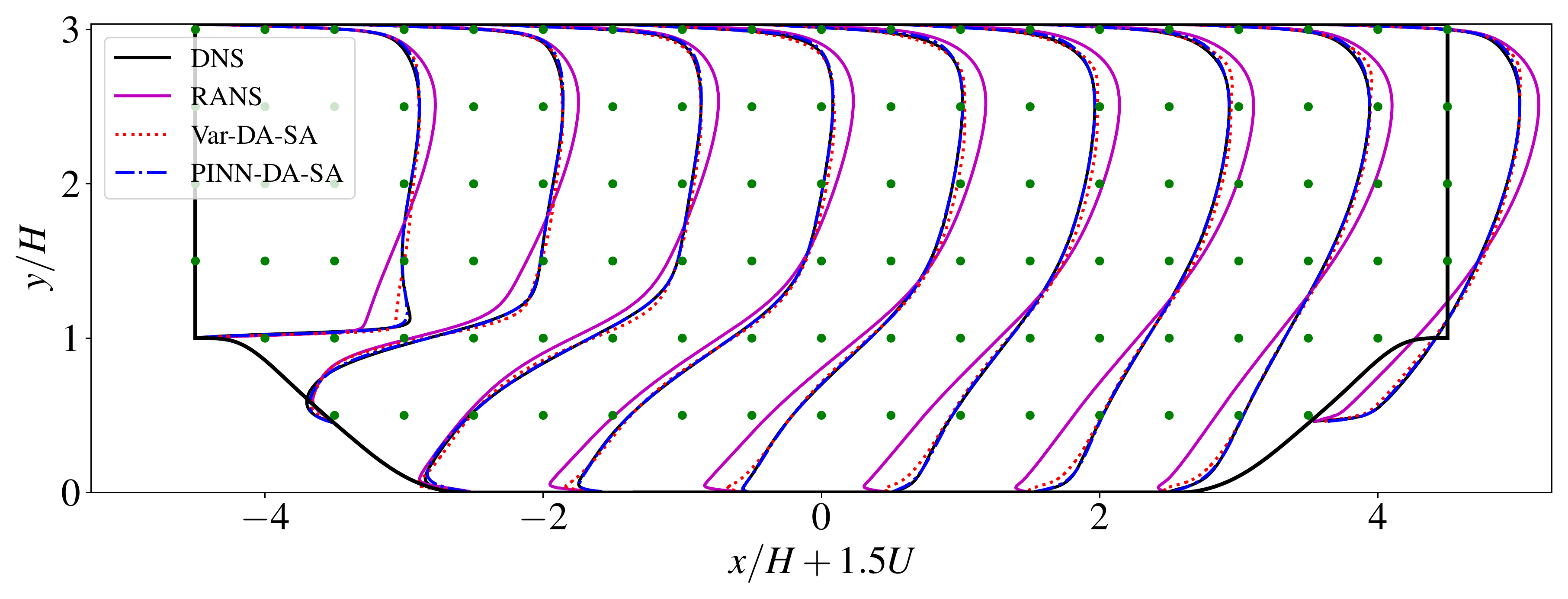}
  \caption{Mean flow reconstruction of streamwise velocity profiles for DNS, RANS-SA, PINN-DA-SA and variational-DA-SA.}
  \label{fig:PINNSA:profile}
\end{figure}

The key feature seen in the comparison between variational-DA-SA and PINN-DA-SA is that the latter has demonstrably lower error across all regions of flow. The highest reduction in error in the PINN-DA-SA is observed in the region between hills ($y/H \leq 1.0$), specifically the recirculation and separation. Whilst all approaches have topologically similar error fields (such as the initial separation), the PINN-DA-SA demonstrates a vastly reduced error. At data points, the reconstruction error is much lower for PINN-DA approaches than the variational-DA-SA. Figure \ref{fig:conv} shows that the PINN-DA's have measurement error of the order of $10^{-7}$ whilst the variational-DA-SA is $10^{-5}$. The relative performance of the variational-DA-SA and PINN-DA-SA compared with both RANS-SA and the DNS flow can be summarised in the velocity profiles in Figure \ref{fig:PINNSA:profile}. Whilst both data assimilation methods represent a significant improvement over a RANS-SA simulation, the PINN-DA-SA is almost identical to the DNS results, whilst small discrepancies are still more apparent for the variational-DA-SA.

Comparing both PINN-DA-SA and variational-DA-SA to PINN-DA-Baseline show that application of SA has been successfully used to reduce error for near-wall gradients, as a result of the additional physical description provided by the turbulence model. The specific choice of turbulence model (SA) is also a contributing factor, given the aforementioned good performance in boundary layers with adverse pressure gradients. 

The lower reconstruction error of the PINN-DA-SA approach compared to the variational-DA-SA approach has been confirmed also for the laminar case, in the absence of the SA model. The test case for this was the unsteady cylinder flow at low Reynolds numbers. Appendix \ref{sec:app:laminar} includes details on the laminar flow case, followed by a comparison of results. 

\begin{figure}[t]
  \centering
  \begin{tabular}[b]{clclc}
    & & \hspace*{0.8cm}$f_{1}$ & & $f_{2}$ \\
    & \small (a) & & \small (b) & \\
    \rotatebox{90}{\hspace*{1.2cm}PINN-DA-Baseline} & & \includegraphics[height=143pt,trim={0.4cm 2.0cm 3.9cm 0.3cm},clip]{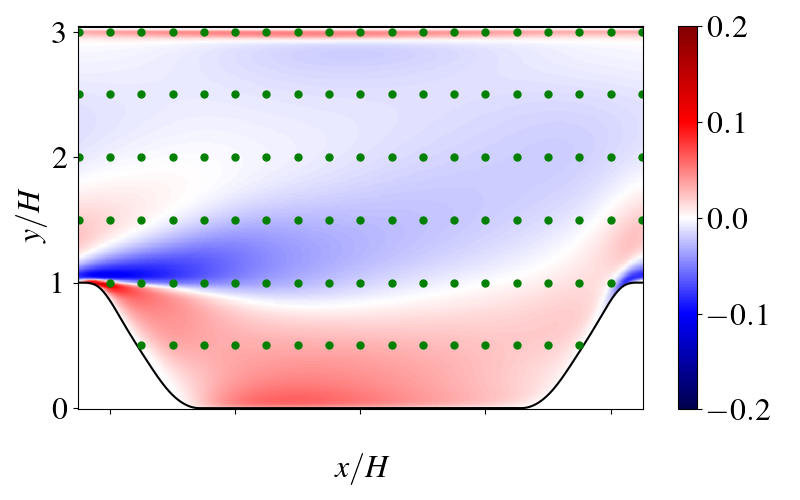} & & \includegraphics[height=143pt,trim={1.9cm 2.0cm 3.8cm 0.3cm},clip]{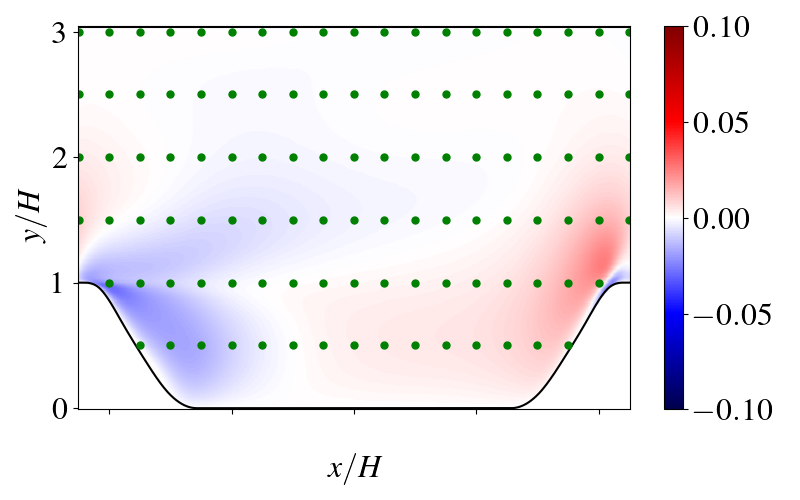} \\
    & \small (c) & & \small (d) & \\  
    \rotatebox{90}{\hspace*{1.6cm}PINN-DA-SA} & & \includegraphics[height=143pt,trim={0.4cm 2.0cm 3.9cm 0.3cm},clip]{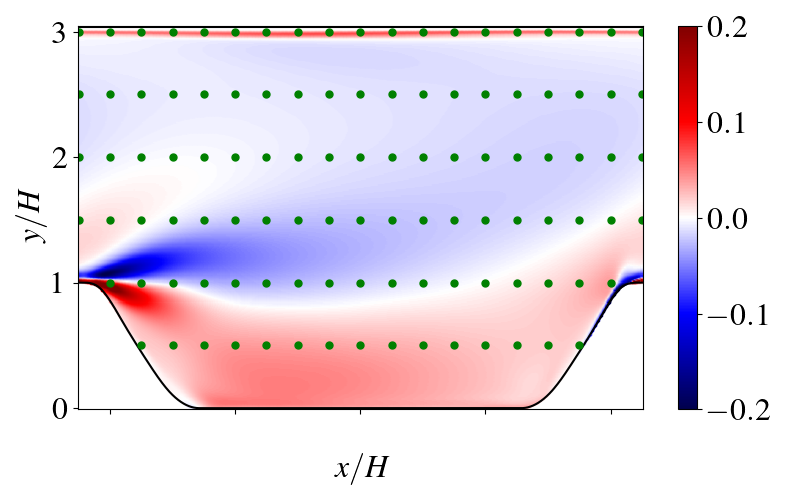} & & \includegraphics[height=143pt,trim={1.9cm 2.0cm 3.8cm 0.3cm},clip]{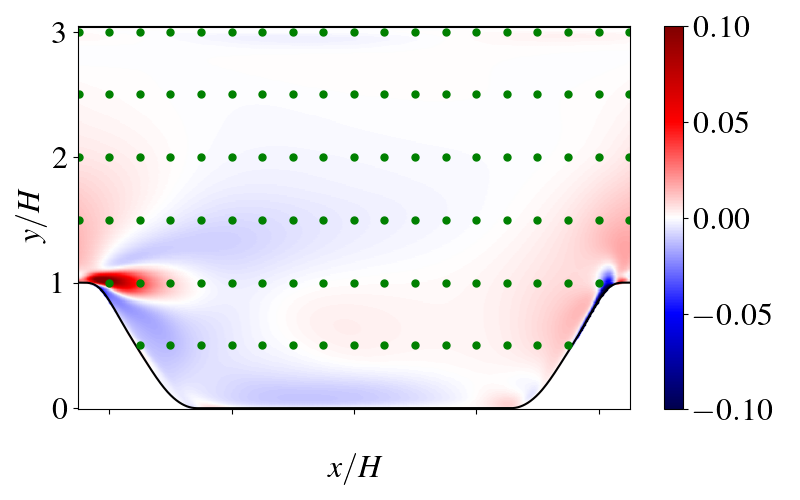} \\
    & \small (e) & & \small (f) & \\
    \rotatebox{90}{\hspace*{1.5cm}Variational-DA-SA} & & \includegraphics[height=163pt,trim={0.4cm 0.5cm 3.9cm 0.3cm},clip]{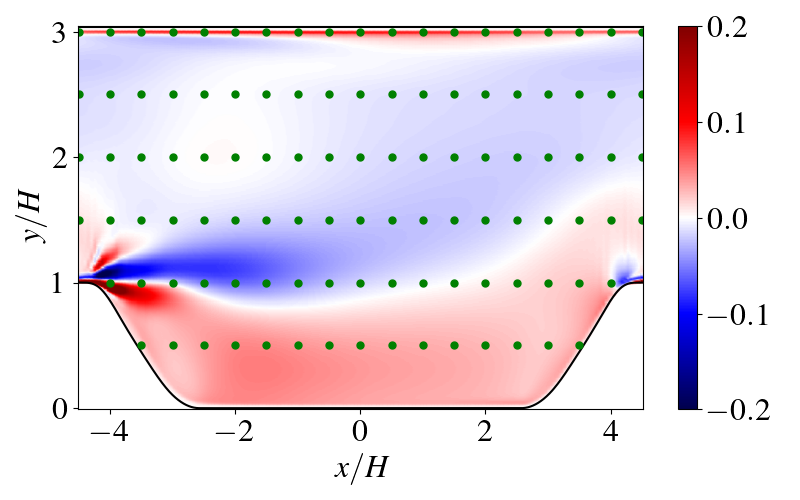} & & \includegraphics[height=163pt,trim={1.9cm 0.5cm 3.9cm 0.3cm},clip]{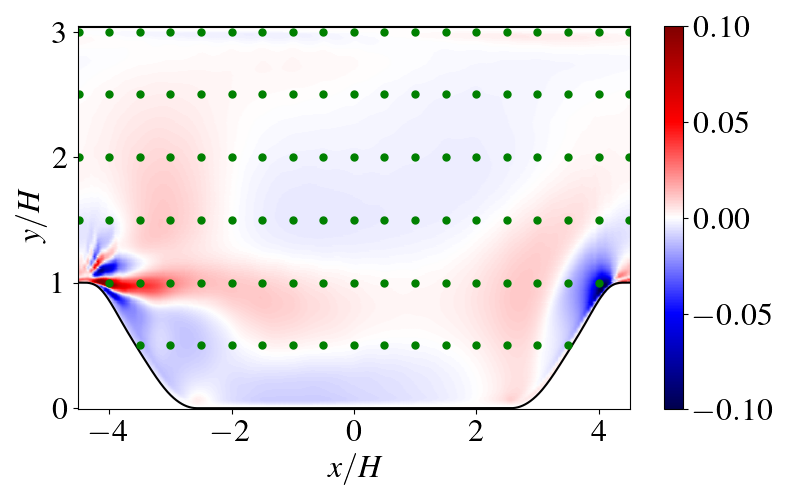} \\
     & & \hspace*{0.80cm}\includegraphics[width=0.42\textwidth,trim={1.1cm 0.7cm 0.4cm 10.1cm},clip]{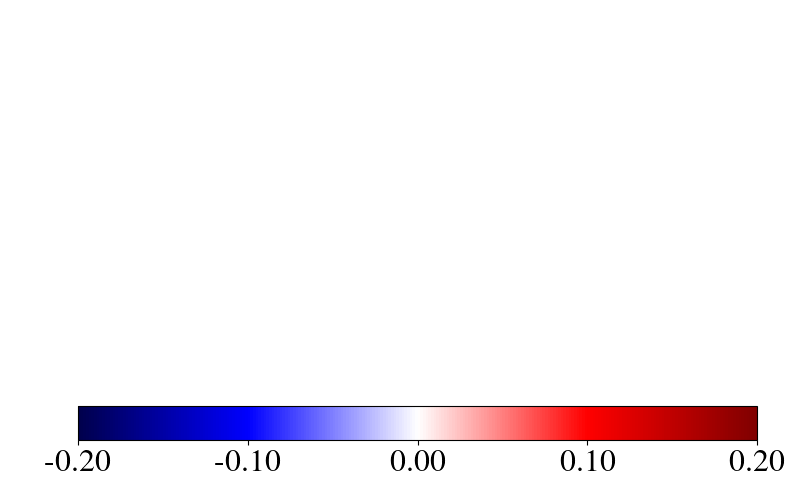} & & \hspace*{0.05cm}\includegraphics[width=0.42\textwidth,trim={1.1cm 0.7cm 0.4cm 10.1cm},clip]{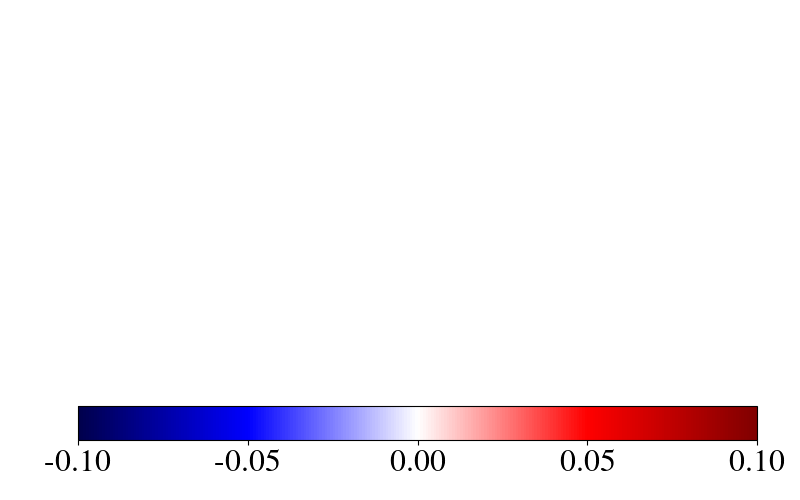}
    \label{fig:fv:PINNSA}
  \end{tabular}
  \caption{A breakdown of the Reynolds forcing (excluding the potential field) $f_1$ (left) and $f_2$ (right) from the different methodologies: $f_{i} = \frac{\partial (2\nu_{t} S_{ij})}{\partial x_j} + f_{s,i}$. For PINN-DA-Baseline, $\nu_{t} = 0$. These are (a,b) - PINN-DA-Baseline, (c,d) - PINN-DA-SA, (e,f) - Variational-DA-SA.}
  \label{fig:forcing}
\end{figure}

\begin{figure}[t]
  \centering
  \begin{tabular}[b]{lclc}
    \small (a) & \hspace*{0.9cm}DNS & \small (b) & PINN-DA-Baseline \\
    & \includegraphics[height=150pt,trim={0.4cm 2.0cm 3.9cm 0.3cm},clip]{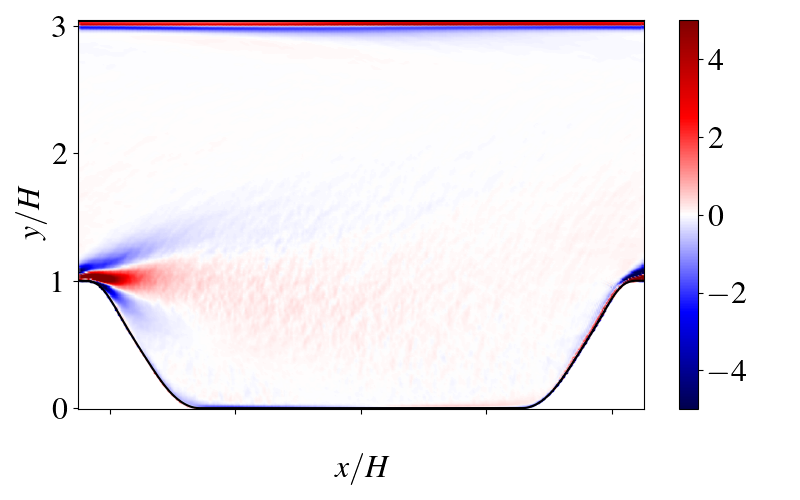} & & \includegraphics[height=150pt,trim={1.9cm 2.0cm 3.9cm 0.3cm},clip]{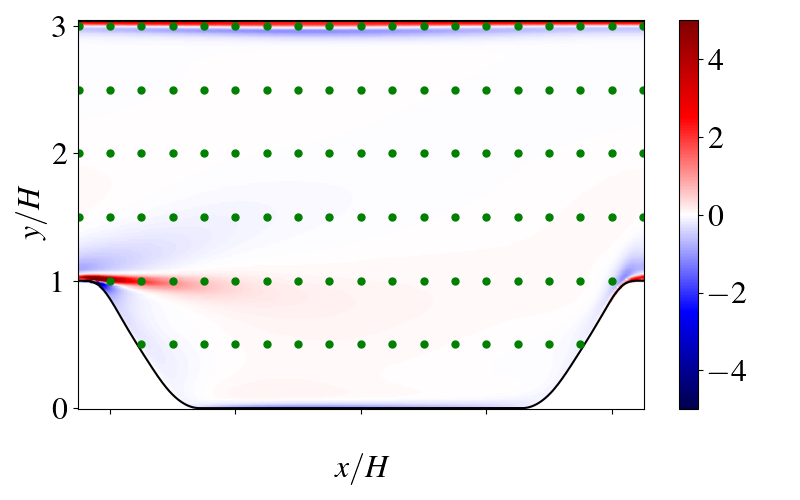} \\
    \small (c) & \hspace*{0.8cm}PINN-DA-SA & \small (d) & Variational-DA-SA\\
    & \includegraphics[height=170pt,trim={0.4cm 0.5cm 3.9cm 0.3cm},clip]{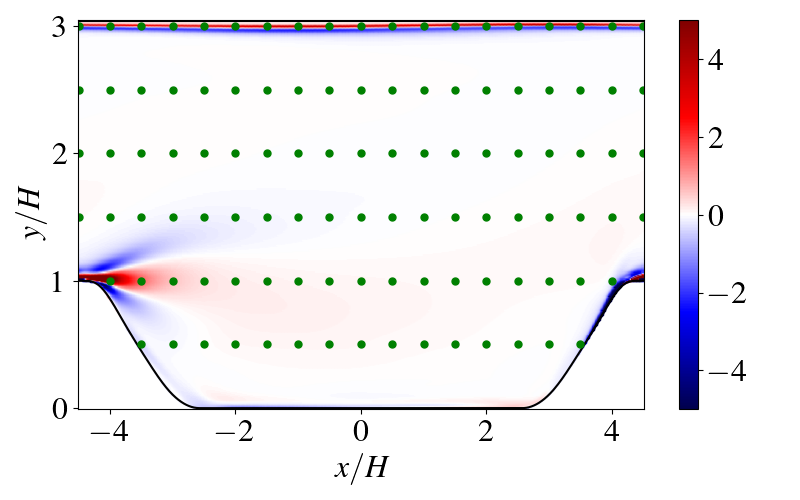} & & \includegraphics[height=170pt,trim={1.9cm 0.45cm 3.9cm 0.3cm},clip]{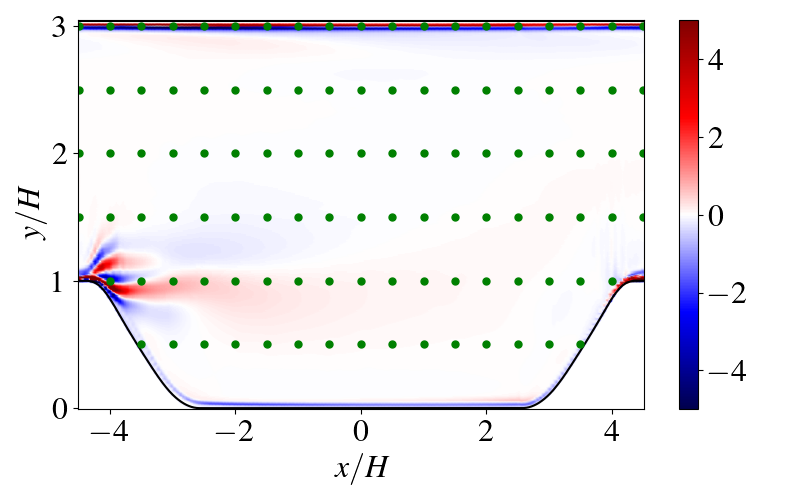}
    \label{fig:curlf:DNS}
  \end{tabular} \\
  \hspace*{1.5cm}\includegraphics[width=0.88\textwidth,trim={1.4cm 0.48cm 0.3cm 10.0cm},clip]{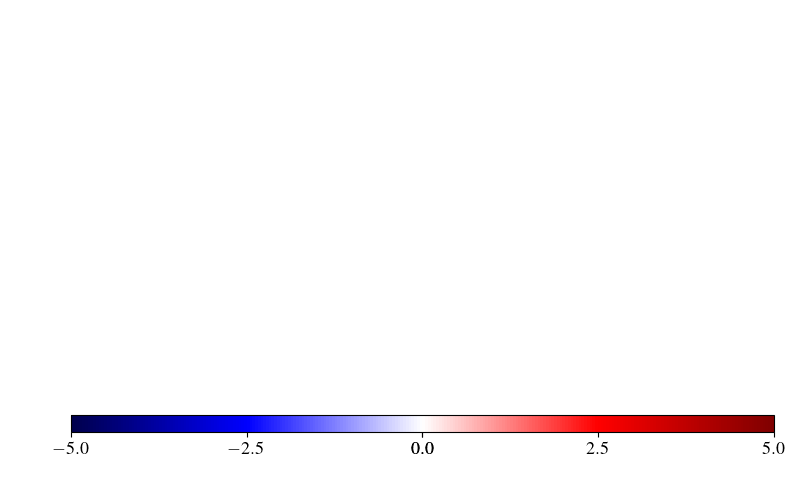}
  \caption{$\nabla\times \mathbf{f}$. Comparison of curl(f) between the ground truth (DNS) and the different data assimilation techniques presented in this paper. (a) DNS, (b) PINN-DA-Baseline, (c) PINN-DA-SA, (d) Variational-DA-SA.}
  \label{fig:curlf}
\end{figure}

\subsubsection{Forcing fields}
The inferred forcing given by Eq. \eqref{eq:eddy:decomp} consists of solenoidal and modelled eddy viscosity (SA contribution) components. For PINN-DA-Baseline, the modelled eddy viscosity component is zero. A further analysis of the forcing terms allows isolation of the effect of SA in the accurate reconstruction of the mean flow field.     

The PINN-DA-Baseline and PINN-DA-SA forcing fields shown in Figure~\ref{fig:forcing} are qualitatively similar in the uniform flow region between $y/H = 1.5$ and $y/H = 2.5$, where turbulent production is low. However, the regions of the flow where PINN-DA-Baseline shows high levels of error (as seen in Figure \ref{fig:PH:comp}(d)) result in larger forcing changes in the PINN-DA-SA solution. Introduction of a modelled component (via SA) has led to significant changes in the corrective forcing, most notably at the separation but also near the walls. These changes underpin the differences in reconstruction error between PINN-DA-Baseline and PINN-DA-SA. Constraining the solution with additional physical definition (via SA) has provided additional modelled forcing in regions where data-assimilation methods with the under-determined RANS had the highest errors. Whilst the PINN-DA-Baseline had converged to a poor solution, the PINN-DA-SA has reduced the \enquote{amount} of under-determined forcing resulting in lower reconstruction errors. However, the addition of the corrective term, which is still unclosed, has allowed the data-assimilation approach to correct the SA model. 

The PINN-DA-SA forcing is different to the variational-DA-SA forcing, in accordance with the differences in mean flow reconstruction. In general, the PINN-DA-SA forcing is closer to the variational-DA-SA forcing than the PINN-DA-Baseline forcing. However, there are several differences in forcing, appearing in regions with the largest differences in reconstruction. This is most apparent in the significantly less smooth forcing around the separation in the variational-DA-SA forcing compared to PINN-DA-SA. The variational-DA-SA also exhibits a high forcing concentration in $f_{2}$ forcing along the rear hill, centred around a datapoint. This datapoint (in the $\Delta L = 0.5$ case) has shown to be particularly sensitive. Whilst the gradient regularisation procedure employed in the variational approach ensures that each update of forcing is smooth, it does not guarantee that the accumulation of updates and thus the final forcing will also be smooth. Whilst other regularisation strategies were investigated (see Appendix \ref{sec:app:reg:var}), they led to poorer reconstructed mean velocity fields.

To evaluate the accuracy of the data assimilation methods, the forcing is also compared with the true DNS forcing. As the potential forcing field is absorbed into the pressure term and thus is inseparable, we instead compare the curl of forcing, to remove the contribution of the potential forcing (since $\mathbf{\nabla\times \nabla}\phi) = 0$).
\begin{equation}
   (\mathbf{\nabla\times f})_{3} = \frac{\partial f_{s,2}}{\partial x} - \frac{\partial f_{s,1}}{\partial y} + \left (\frac{\partial^2}{\partial x^2} - \frac{\partial^2}{\partial y^2}\right )2\nu_{t}S_{12} + \frac{\partial^2}{\partial x \partial y} \left (2\nu_{t}(S_{22} - S_{11})\right ).
    \label{eq:RANS:curlf:DA}
\end{equation}
A direct comparison can be performed then against the curl of the DNS forcing
\begin{equation}
    (\mathbf{\nabla\times f})_{3} = \frac{\partial f_{2}}{\partial x} - \frac{\partial f_{1}}{\partial y} = \left (\frac{\partial^2}{\partial x^2} - \frac{\partial^2}{\partial y^2}\right )\overline{{u}'{v}'} + \frac{\partial^2}{\partial x \partial y} \left (\overline{{v}'{v}'} - \overline{{u}'{u}'}\right ).
    \label{eq:RANS:curlf}
\end{equation}

The curl of the inferred forcing for all three reconstruction methods employed here is shown in Figure \ref{fig:curlf}. Best agreement with the DNS is achieved for PINN-DA-SA, which aligns with the most accurate mean flow reconstruction. The PINN-DA-Baseline differs from DNS in several key regions - the initial separation, the recirculation, the rearward hill apex and along the walls. 
Differences between the variational-DA-SA and PINN-DA-SA forcing (and curl) correlate with differences in the error fields. These are most apparent at the separation region and along the upper and lower walls. 

\subsubsection{Data Assimilation with varying data resolution}
\begin{figure}[t]
    \centering
    \includegraphics[width=0.65\textwidth,trim={0.5cm 0.5cm 0.3cm 0.3cm},clip]{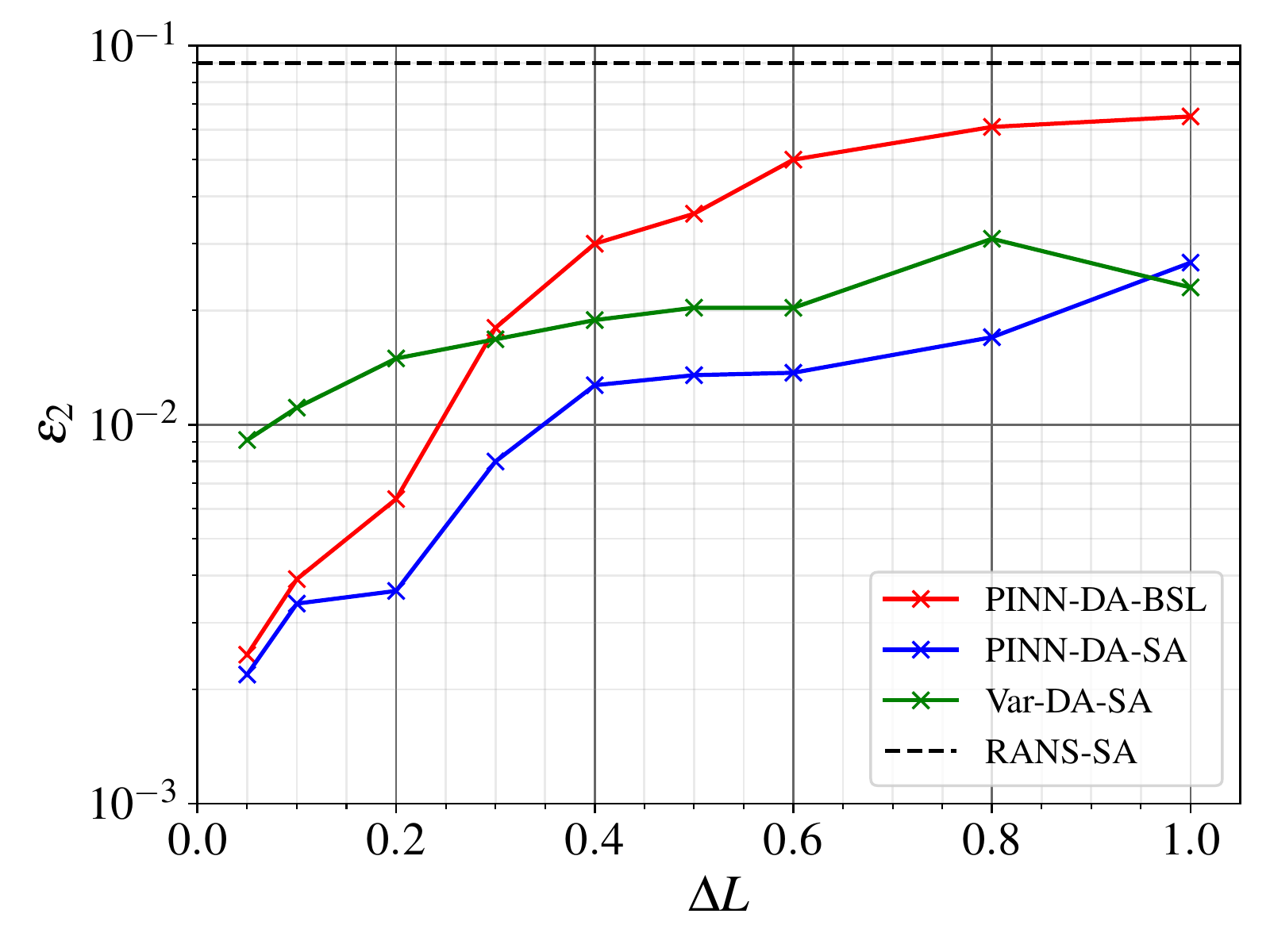}
    \caption{Volume-weighted $L_2$ mean velocity reconstruction error (as defined in Eq. \eqref{eq:L2}) versus data resolution for the three reconstruction methodologies (solid lines with symbols). All three approaches improve the RANS-SA predictions (dashed line).}
    \label{fig:PH:L2}
\end{figure}

Here, we evaluate the dependence of the reconstruction error on the data resolution of the available high-fidelity measurements. In Figure~\ref{fig:PH:L2}, we compare variational-DA-SA, PINN-DA-Baseline and PINN-DA-SA approaches, for a range of data-resolutions, ranging from $\Delta L=0.05$ to $\Delta L=1.0$. The reconstruction error increases monotonically as the (equidistant/square) spacing between data points increases. All data assimilation methods improve the RANS-SA solution, irrespective of resolution. For data spacing $\Delta L < 1$, the turbulence model augmented PINN-DA-SA method achieves the lowest reconstruction error against all the other methods. Additionally, whilst the $\Delta L=0.8$ has not converged as well for the variational-DA-SA approach, the reconstruction error remains close to the neighbouring points.

Comparing the PINN-DA-Baseline to the PINN-DA-SA results shows the significant improvement achieved through the turbulence model augmentation. The gap between the two can mostly be attributed to the error near the walls and in the separation. Additionally, the rate of error increase with respect to data resolution is steeper for PINN-DA-Baseline. This suggests adding a modelled (closed) forcing component (i.e. PINN-DA-SA) reduces the effect coarsening data resolution has on reconstruction accuracy compared to PINN-DA-Baseline. As data resolution becomes finer, the two PINN-DA approaches converge. This is expected - as more data is provided, there are more points at which the PINN-DA-Baseline can close the solution and thus the advantage the turbulence model augmentation provides is diminished.

Comparing the PINN-DA-SA versus the variational-DA-SA solution shows that, for most of the resolution range, the PINN-DA-SA also demonstrates consistently lower error at equivalent resolutions. In fact, the PINN-DA-SA at $\Delta L = 0.8$ matches the reconstruction accuracy of the variational-DA-SA at $\Delta L = 0.3$. However, at the extreme end of coarse data ($\Delta L = 1.0$), whilst the PINN-DA-SA is still a significant improvement over the PINN-DA-Baseline (and also RANS-SA), the variational-DA-SA approach has a lower reconstruction error. This suggests that the PINNs have a greater dependence on data for good performance. As data resolution reduces (spacing increases), the optimisation problem converges to solving the direct RANS-SA problem, for which the numerical framework within the variational-DA-SA method is more robust i.e. PINNs are less effective forward solvers. Algorithmic improvements for PINNs, including adaptive weighting algorithms, can potentially provide further improvements of the PINN approach (as demonstrated by ~\citep{Wang:2022}), which is currently under investigation. On the opposite end, for fine data resolutions, the variational-DA-SA is less accurate than the PINN-DA-Baseline.

One may attribute the difference between the reconstruction accuracy of the PINN-DA-SA and variational-DA-SA to the differences in the way the RANS equations are enforced in the data assimilation procedure. Concerning the variational-DA-SA approach, it may be first noticed that the latter solves for only the weak form of the RANS equations, following the finite-element approach. In addition, as mentioned in Section \ref{sec:var_setup}, stabilisation terms are added in the governing equations. Then there is the process of spatial discretisation through, among others, the choice of shape functions, which will determine the order of spatial accuracy. In contrast, PINNs enforce the RANS equations in their original, strong formulation, and in a continuous way. Moreover, the effect of applying discrete pointwise forcing at measurement locations poses a greater challenge for the variational approach than PINNs and thus the method of regularisation has a greater impact on reconstruction performance. Appendix \ref{sec:app:reg:var} contains analysis of regularisation choice. By avoiding both these sources of error, PINN-DA-SA achieves a lower reconstruction error than the variational-DA-SA. This is evidenced by the reconstruction error for finer data resolutions. Even as data resolution increases, the aforementioned discretisation error remains. Consequently, the reconstruction error of the variational-DA-SA reduces at a much slower rate than both PINN-DAs.

\section{\label{sec:conclusion}Conclusion}
Three approaches to turbulent mean flow reconstruction from sparse data measurements have been presented and compared. Firstly, it was shown that for all methods, inferring the closure of the underdetermined RANS equations or the corrective closure of the RANS-SA equations by assimilating high-fidelity sparse data improved the mean flow reconstruction compared to RANS-SA predictions. 

Secondly, the use of Spalart-Allmaras turbulence model, as used in ~\citep{Franceschini:2020} (in the variational-DA-SA approach) was proposed for use in PINNs, particularly in the context of turbulent flows. The SA turbulence model augmented PINN (PINN-DA-SA) proved to be the most accurate flow reconstruction method followed by the equivalent approach albeit with a variational formulation. Comparing the PINN-DA-Baseline (PINN without turbulence SA model) with PINN-DA-SA showed the isolated effect of turbulence model augmentation. The PINN-DA-SA performs significantly better in regions with complex flow features, such as near walls and at separation. Injection of additional physical constraints in the form of a turbulence model provides an additional modelled component to the underdetermined Reynolds forcing, reducing the complexity of the optimisation and narrowing the solution space for the optimisation. A corrective forcing term is then tuned to correct the rest of the closure.

Thirdly, the variational-DA-SA approach, whilst an improvement over the PINN-DA-Baseline, has higher errors compared to the PINN-DA-SA, most notably off the separation. These observations for mean flow reconstruction accuracy are also associated with equivalent error in Reynolds forcing. The difference in variational-DA-SA and PINN-DA-SA can be attributed to the effect of discretisation, which is avoided by PINNs. These observations are consistent across different data resolutions.

Given that PINNs is a growing field of research, there are potentially many avenues to improve the reconstruction capability further. The most apparent example is hyper-parameter tuning, such as loss weighting. In this study, a static weighting method was used with manual iterations to tune them. This is a laborious exercise, particularly as systems become more complex with more boundary conditions and less uniform topologies. Adaptive weighting algorithms~\citep{Wang:2021,Wang:2022,Van:2022} or novel architectures, such as Competitive PINNs~\citep{Zeng:2022}, would both simplify this problem and also generalise the process to selecting weights. This has been demonstrated in Appendix \ref{sec:app:hyper}.

\

\noindent {\bf Code and data availability:} the code and data to run the PINN cases presented here are available at \url{https://github.com/RigasLab/PINN_SA}.
\appendix
\section{Spalart-Allmaras equations\label{sec:app:SA}}
The full definition of the SA transport equations, as mentioned in Equation \eqref{eq:RANS:SA:helm}, is described here.
\subsection{Formulation of Spalart-Allmaras equations\label{sec:app:SA:BSL}}
The Spalart-Allmaras  equations are solved for $\tilde{\nu}$. Then, the eddy viscosity, $\nu_{t}$, is obtained as
\begin{equation}
    \nu_{t} =
    \begin{cases}
        0 & \text{ if } \tilde{\nu} < 0, \\
        f_{v1}\tilde{\nu} & \text{ if } \tilde{\nu} \geq 0,
    \end{cases}
\end{equation}
where
\begin{equation}
    f_{v1} = \frac{\chi^3}{\chi^3 + c_{v1}^3}, \hspace{10pt}\chi = \frac{\tilde{\nu}}{\nu}.
\end{equation}
The SA transport equation takes the form
\begin{equation}
    U_{j}\frac{\partial \tilde{\nu}}{\partial x_j} - S_{p} - S_{diff} - S_{c} - S_{d} = 0,
\end{equation}
where $S_{p}, S_{diff}, S_{c}, S_{d}$ are production, diffusion, cross-diffusion and destruction terms respectively. These are defined as
\begin{equation}
    \begin{array}{cc}
        S_{p} = 
        \begin{cases}
            c_{b1}S\tilde{\nu}g_{n} & \text{ if } \tilde{\nu} < 0, \\
            c_{b1}\tilde{S}\tilde{\nu} & \text{ if } \tilde{\nu} \geq 0,
        \end{cases} &
        S_{d} = 
        \begin{cases}
            c_{w1}\frac{\tilde{\nu}^2}{d^2} & \text{ if } \tilde{\nu} < 0, \\
            -c_{w1}f_{w}\frac{\tilde{\nu}^2}{d^2} & \text{ if } \tilde{\nu} \geq 0,
        \end{cases}\\
        S_{diff} = \frac{1}{\sigma}\nabla\cdot(\eta\nabla\tilde{\nu})
        = \frac{1}{\sigma}\left( \eta\frac{\partial ^{2} \tilde{\nu}}{\partial x_j^{2}} + \frac{\partial \eta}{\partial x_j}\frac{\partial \tilde{\nu}}{\partial x_j}\right), & \hspace{15pt}
        S_{c} = \frac{c_{b2}}{\sigma}\left \| \nabla\tilde{\nu} \right \|^{2}
        = \frac{c_{b2}}{\sigma} \left( \frac{\partial \tilde{\nu}}{\partial x_j} \right)^{2}.
    \end{array}
\end{equation}
where the auxillary functions and constants are defined as
\begin{equation}
    \begin{array}{ccc}
        \eta = 
        \begin{cases}
            \nu(1 + \chi + \frac{1}{2}\chi^2) & \text{ if } \tilde{\nu} < 0, \\
            \nu(1 + \chi) & \text{ if } \tilde{\nu} \geq 0,
        \end{cases} &
        {S}' = \left \| \nabla\times\mathbf{U} \right \|, &
        {S} = \sqrt{{S}'^2 + M^2} - M, \\
        \overline{S} = \frac{\tilde{\nu}f_{v2}}{\kappa^2 d^2}, &
        f_{v2} = 1 - \frac{\chi}{1 + \chi f_{v1}}, & 
        \tilde{S} = max(10^{-10}, S + \overline{S}), \\
        {r}' = \frac{\tilde{\nu}}{\tilde{S}\kappa^2 d^2}, &
        r = 
        \begin{cases}
            {r}' & \text{ if } 0 \geq {r}' \geq 10, \\
            0 & \text{ otherwise, }
        \end{cases} &
        g = r + c_{w2}(r^6 - r), \\
        f_{w} = g\left(\frac{1 + c_{w3}^6}{g^6 + c_{w3}^6} \right)^{\frac{1}{6}}, & &
        g_{n} = 1 - \frac{1000\chi^2}{1 + \chi^2}.
    \end{array}
\end{equation}
The constants in the SA transport equations are
$c_{v1} = 7.1$, $c_{v2} = 0.7$, $c_{v1} = 0.9$, 
$\kappa = 0.41$, $\sigma = \frac{2}{3}$, 
$c_{b1} = 0.1355$, $c_{b2} = 0.622$,  
$c_{w1} = \frac{c_{b1}}{\kappa^2} + \frac{1 + c_{b2}}{\sigma}$, $c_{w2} = 0.3$, $c_{w3} = 2$, $M=10^{-5}$.

\subsection{Wall-Distance multiplied Spalart-Allmaras Equations\label{sec:app:SA:Mod}}
To remove singularities, which cause problems for the PINN-DA-SA formulation at points evaluated at (or near) the wall (as discussed in Section \ref{sec:numerics:PINN:mod}), the modified SA equations take the form
\begin{equation}
    d^{2}\left ( U_{j}\frac{\partial \tilde{\nu}}{\partial x_j}  - S_{diff} - S_{c}\right ) - S_{p}^{N} - S_{d}^{N} = 0,
\end{equation}
where $S_{p}^{N},S_{d}^{N}$ are new modified production and destruction terms, respectively. These are defined as
\begin{equation}
    \begin{array}{cc}
        S_{p}^{N} = d^{2}S_{p} = 
        \begin{cases}
            c_{b1}S^{N}\tilde{\nu}g_{n} & \text{ if } \tilde{\nu} < 0, \\
            c_{b1}\tilde{S}^{N}\tilde{\nu} & \text{ if } \tilde{\nu} \geq 0,
        \end{cases} & \hspace{15pt}
        S_{d}^{N} = d^{2}S_{d} = 
        \begin{cases}
            c_{w1}\tilde{\nu}^2 & \text{ if } \tilde{\nu} < 0, \\
            -c_{w1}f_{w}\tilde{\nu}^2 & \text{ if } \tilde{\nu} \geq 0.
        \end{cases}
    \end{array}
\end{equation}
The modified auxiliary functions are thus
\begin{equation}
    \begin{array}{cc}
        S^{N} = d^{2}S = d^{2} \left (\sqrt{{S}'^2 + M^2} - M\right ), &
        \overline{S}^{N} = d^{2}\overline{S} = \frac{\tilde{\nu}f_{v2}}{\kappa^2}, \\
        \tilde{S}^{N} = d^{2}\tilde{S} = max(10^{-10}, d^{2}\left (S + \overline{S}\right )) = max(10^{-10}, S^{N} + \overline{S}^{N}), &
        {r}' = \frac{\tilde{\nu}}{\tilde{S}^{N}\kappa^2}.
    \end{array}
\end{equation}

\section{\label{sec:app:reg}Effect of regularisation on data assimilation techniques}
As mentioned in Section \ref{sec:methodology:data}, the forcing decomposition between modelled and corrective parts is not unique and thus the regularisation of corrective forcing is critical. This section will show the effect of regularisation for both PINN-DA and variational approaches and highlight its importance.

\subsection{\label{sec:app:reg:PINN}Effect of solenoidal forcing regularisation on PINNs}
\begin{figure}[t]
  \centering
  \includegraphics[width=0.8\textwidth,trim={0.1cm 0.1cm 0.1cm 0.1cm},clip]{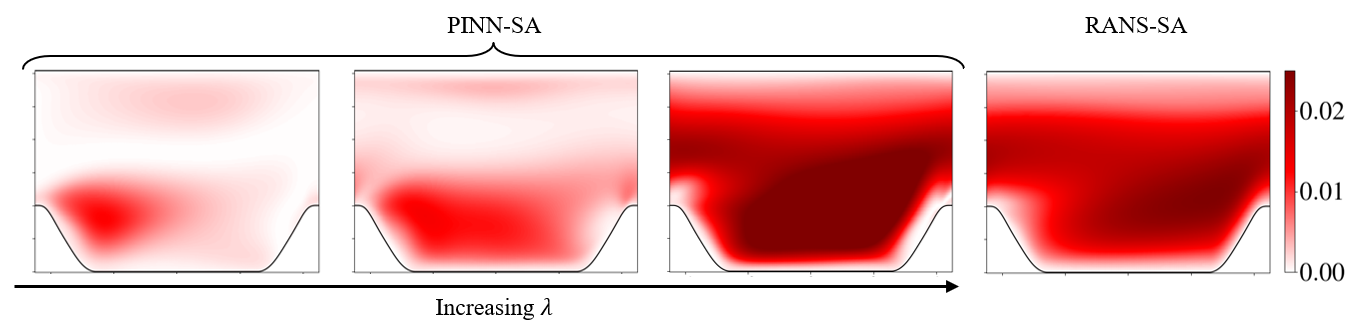}
  \caption{Effect of regularisation parameter (solenoidal forcing magnitude penalisation), $\lambda^R$ on PINN-SA solution for $\tilde{\nu}$. $\tilde{\nu}$ for increasing $\lambda^R$ compared with the RANS-SA $\tilde{\nu}$ shows that increasing regularisation causes eddy viscosity to converge towards the RANS solution. The presented values of $\lambda^R$ are $10^{-8}$, $10^{-4}$, $10^{0}$ left to right}
  \label{fig:eddy:reg}
\end{figure}

For PINN-DA-SA, an $L_2$ penalisation of the solenoidal forcing magnitude was applied. During PINN-DA-SA hyper-parameter tuning, it was found that the final solution was strongly influenced by the regularisation of solenoidal forcing. Clearly as regularisation weight, $\lambda^R$, increases, the magnitude of solenoidal forcing is penalised more and thus there is a greater dependence on the modelled eddy viscosity forcing component. In effect, the greater the value of $\lambda$, the closer the governing equations tend towards a pure RANS-SA problem. However, for the PINN-DA, enforcement of data loss at measurement points results in differences. This is demonstrated in Figure \ref{fig:eddy:reg}. As solenoidal forcing magnitude is penalised more and more, the solution tends towards RANS-SA. However, this is an imperfect solution as the PINN-DA-SA is attempting to fit high fidelity (mean) DNS data to the RANS-SA equations. This is most notable at the point around the rear hill apex. As a result, the residual PDE loss begins to increase at very high $\lambda^R$. On the other hand, without penalisation ($\lambda^R = 0$), the PINN-DA-SA solution tends towards the PINN-DA-Baseline solution with a negligible contribution from the eddy viscosity component ($\tilde{\nu} \approx 0$). 

\begin{figure}[t]
  \centering
  \includegraphics[height=150pt,trim={0.2cm 0.5cm 0.3cm 0.2cm},clip]{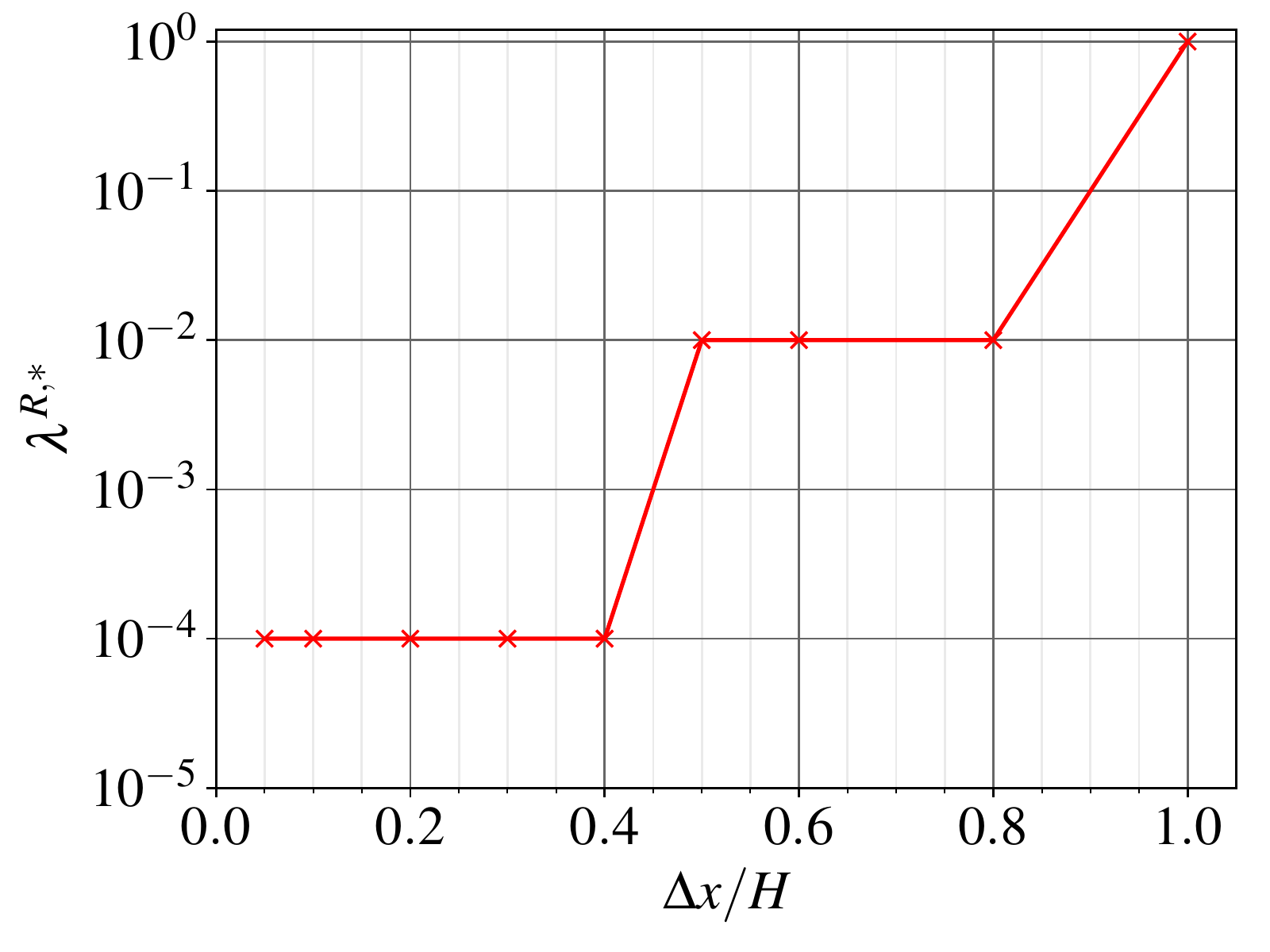}
  \caption{The optimised regularisation parameter, $\lambda^{R,*}$, with respect to data resolution.}
  \label{fig:eddy:regvsres}
\end{figure}

As the data coarsens, and thus the dependence on data reduces, the requirement for a modelled component becomes more important, as the corrective Reynolds forcing can be closed at fewer points using data. Less data means the solution will move closer to a RANS-SA solution. As a result, the coarser the data, the greater the value for the optimal $\lambda^{*}$, as the modelled forcing component becomes a greater component of the forcing. This is seen in Figure \ref{fig:eddy:regvsres}.

\subsection{\label{sec:app:reg:var}Importance of regularisation for variational DA}
For the variational DA, the choice of regularisation (of solenoidal forcing $f_{s,i}$) was more important. Three approaches were trialled: no regularisation; $L_2$ regularisation; $H_1$ regularisation.

\begin{figure}[t]
  \centering
  \begin{tabular}[b]{clclclc}
    & & \hspace*{0.5cm}No Regularisation & & $L_{2}$ Regularisation & & $H_{1}$ Regularisation\hspace*{0.5cm}\\
    & \small (a) & & \small(b) & & \small(c) & \\
    $U_{err}$ & & \parbox[c]{139pt}{\includegraphics[height=89pt,trim={0.5cm 2.4cm 4.25cm 0.3cm},clip]{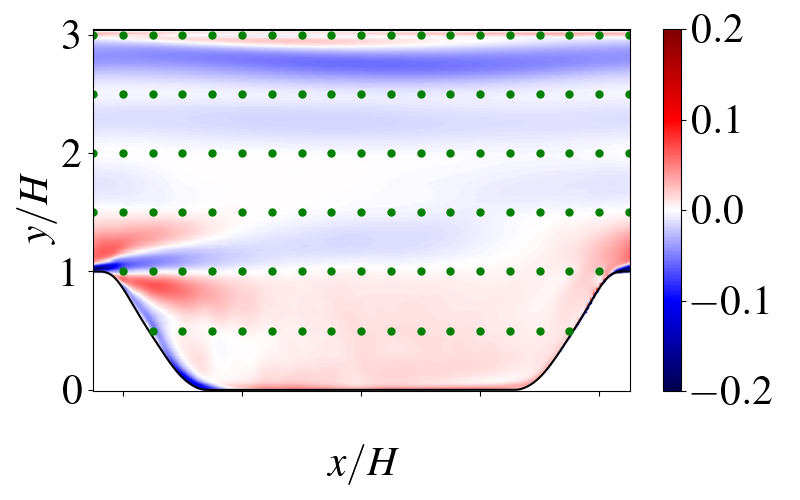}} & & \parbox[c]{124pt}{\includegraphics[height=89pt,trim={2.3cm 2.4cm 4.25cm 0.3cm},clip]{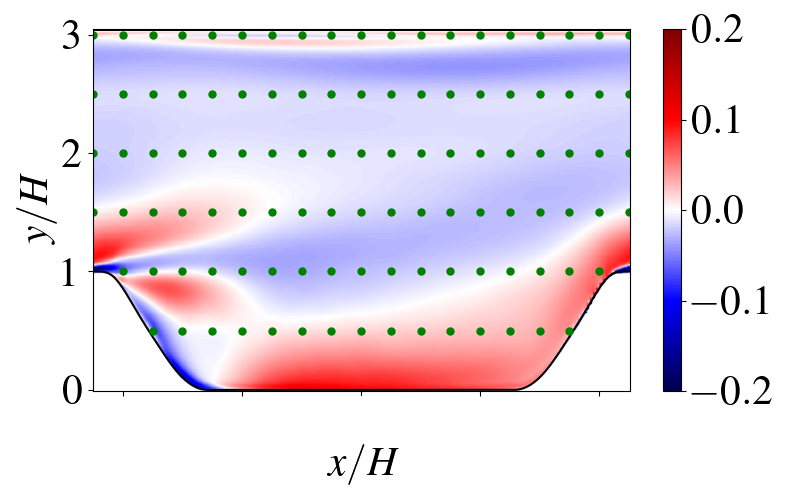}} & & \parbox[c]{154pt}{\includegraphics[height=89pt,trim={2.3cm 2.4cm 0.7cm 0.3cm},clip]{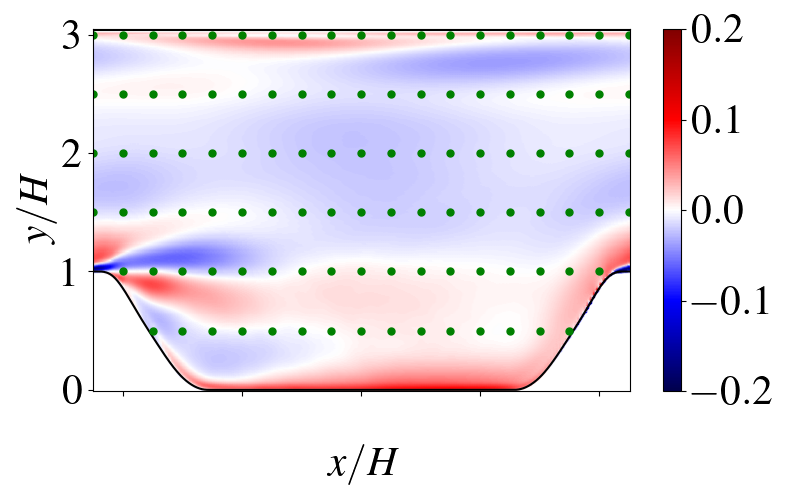}}\\
    & \small (d) & & \small(e) & & \small(f) & \\
    $\left \| f_{s,i} \right \|$ & & \parbox[c]{139pt}{\includegraphics[height=104pt,trim={0.5cm 0.6cm 4.0cm 0.3cm},clip]{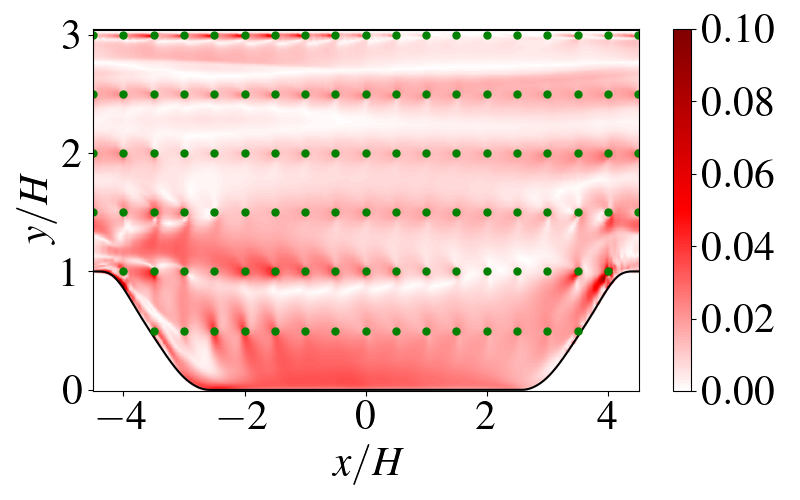}} & & \parbox[c]{124pt}{\includegraphics[height=104pt,trim={2.3cm 0.6cm 4.0cm 0.3cm},clip]{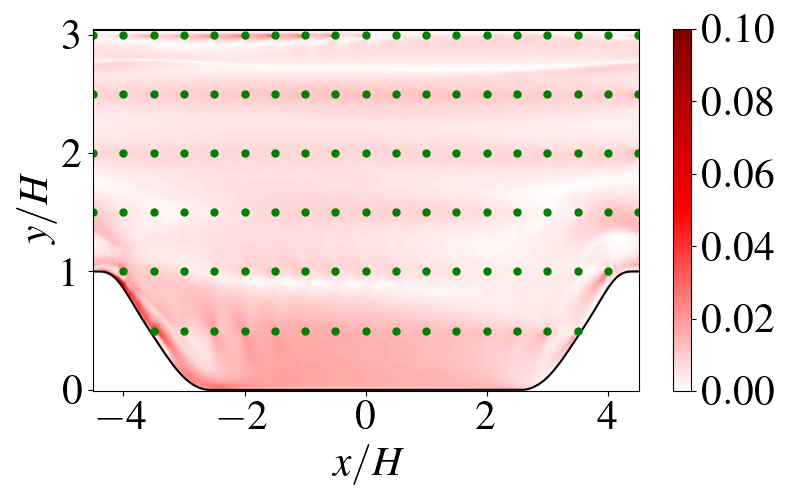}} & & \parbox[c]{153pt}{\includegraphics[height=104pt,trim={2.3cm 0.6cm 0.65cm 0.3cm},clip]{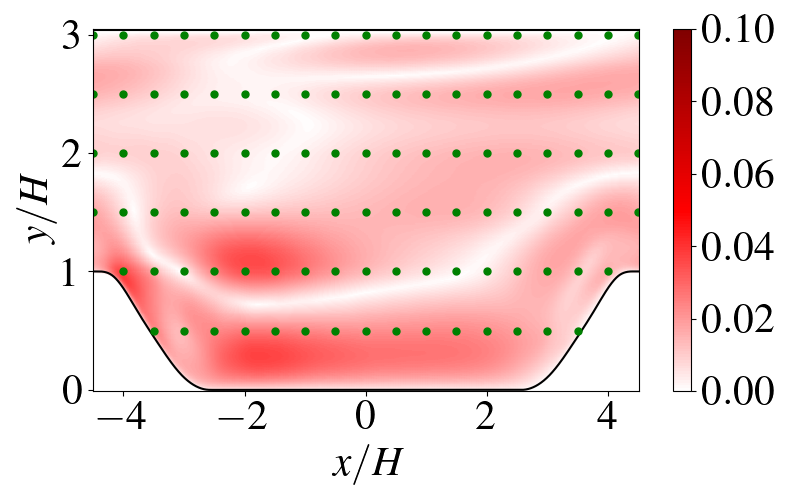}}\\
  \end{tabular}
  \caption{Absolute $U$ error (a,b,c) and corrective solenoidal forcing, $\left \| f_{s,i} \right \|$,(d,e,f) for variational-DA-SA method using different regularisation approaches: No regularisation (left), $L_{2}$ regularisation (middle) and $H_{1}$ regularisation (right).}
  \label{fig:PH:var:reg}
\end{figure}

Firstly, without regularisation, whilst data is well enforced (as seen in Figure \ref{fig:PH:var:reg}(a)) even compared with the final $H_{1}$ based approach, it results in a non-physical solenoidal forcing that appears strongly pointwise, clearly seen in Figure \ref{fig:PH:var:reg}(d), as large corrections are applied at the data locations. The final reconstruction is thus less physical. Additionally, regularisation is needed to enforce convergence to a unique decomposition.

For the $L_{2}$ regularisation approach, error is higher than either $H_{1}$ or no regularisation methods. This is due to the inefficiency of the approach in the variational framework. To achieve sufficient smoothing of the solution, avoiding the problems of the none regularised approach, very large penalisation weighting is needing. However, this comes at the cost of poor data enforcement, as seen in Figure \ref{fig:PH:var:reg}(b). 

For the variational-DA-SA, $H_1$ regularisation was selected to smooth out the solution. As demonstrated, for the variational approach, $H_{1}$ regularisation is a much more suitable approach. It allows much better matching of the data (as compared with $L_2$ regularisation), whilst also giving a more physical solution. As a result this was selected as the final regularisation method.

For PINN-DA-SA, good data enforcement and a smooth non-pointwise solenoidal forcing was always ensured. Regularisation is exclusively used to control the distribution between modelled and corrective forcing. Given its computational simplicity, $L_{2}$ was selected for PINN-DA-SA. This was not the case for variational-DA-SA and regularisation was needed to smooth the solution in addition to the distribution of modelled and corrective forcing.

\section{\label{sec:app:hyper}Hyper-parameter and loss weight selection}
\subsection{Tuning the model with a grid search approach}
To select hyper-parameters (model architecture, activation function and learning rate) and loss weighting, a grid search method was used. Tables \ref{tab:hyp:act} and \ref{tab:hyp:arch}, show the sensitivity of final model loss to hyper-parameters. From these results, the architecture, as discussed in Section \ref{sec:numerics:PINN}, was selected. A similar analysis was completed for the loss weighting. A summary of the key results are shown in Table \ref{tab:weight:grid}. Repeats of the final model were found to converge within $3.0\%$ of the loss presented in Section \ref{sec:res:conv} (after 3 repetitions).
\begin{table}[H]
\caption{Sensitivity of final model loss function, $C$, to activation functions and learning rate on a model with 7 layers and 50 nodes/layer relative to final choice ($\tanh$ and $5\times10^{-4}$ learning rate)}
\label{tab:hyp:act}
\begin{ruledtabular}
\begin{tabular}{ccccc|ccc}
\multicolumn{5}{c|}{Activation Functions}  & \multicolumn{3}{c}{Learning Rate (with $\tanh$ activation)}  \\ \hline
$\tanh$ & ReLU & $\sin$ & ELU & L-LAAF $\tanh$ & $1\times 10^{-3}$ & $5\times 10^{-4}$ & $1\times 10^{-4}$ \\ \hline
- & $+55.4\%$ & $+9.1\%$ & $+69.3\%$ & $+5.1\%$ & $-1.3\%$ & - & $+3.6\%$  \\
\end{tabular}
\end{ruledtabular}
\end{table}
\begin{table}[H]
\caption{Sensitivity on final model loss function, $C$, to architecture parameters (number of layers and nodes) relative to final architecture (7 layers, 50 nodes).}
\label{tab:hyp:arch}
\begin{ruledtabular}
\begin{tabular}{cc|ccccc}
& & \multicolumn{5}{c}{Number of Layers} \\
& & 5 & 6 & 7 & 8 & 9 \\ \hline 
& 25  & $+19.7\%$ & $+15.6\%$ & $+9.9\%$ & $+7.5\%$ & $+5.2\%$ \\
\multirow{1}{*}{Number of Nodes/Layer} & 50  & $+7.4\%$ & $+5.7\%$ & - & $+2.8\%$ & $+2.6\%$ \\
& 75  & $+6.3\%$ & $+5.5\%$ & $+2.8\%$ & $+4.7\%$ & $+6.3\%$ \\
\end{tabular}
\end{ruledtabular}
\end{table}
\begin{table}[H]
\caption{Sensitivity on final model loss function, $C$, to perturbations on weights from Table \ref{tab:weight}}
\label{tab:weight:grid}
\begin{ruledtabular}
\begin{tabular}{cc|cc|cc}
\multicolumn{2}{c|}{$\lambda^P$} & \multicolumn{2}{c|}{$\lambda^D$}  & \multicolumn{2}{c}{$\lambda^B$}    \\
$\times 10$ & $\times 0.1$ & $\times 10$ & $\times 0.1$ & $\times 10$ & $\times 0.1$ \\ \hline
$+4.7\%$    & $+8.6\%$ & $+3.8\%$   & $+9.3\%$  & $+1.6\%$    &  $+7.2\%$   \\
\end{tabular}
\end{ruledtabular}
\end{table}
A $\tanh$ activation was selected due to the lowest value of cost function. Whilst an initial learning rate of $1\times 10^{-3}$ demonstrated a slightly lower final loss value, this model was less repeatable than the chosen learning rate ($5\times 10^{-4}$). Similarly, the model architecture (nodes and layers) was selected for demonstrating the lowest final loss value. Whilst adding more layers and nodes performs similarly, the additional model complexity increases computational cost.

\subsection{Alternative tuning methodologies}
\begin{figure}[t]
  \centering
  \begin{tabular}[b]{lclc}
    & Adaptive Activation Function & & Adaptive Weighting Algorithm \\
    \small (a) & & \small(b) & \\
    & \includegraphics[height=170pt,trim={0.4cm 0.5cm 0.3cm 0.3cm},clip]{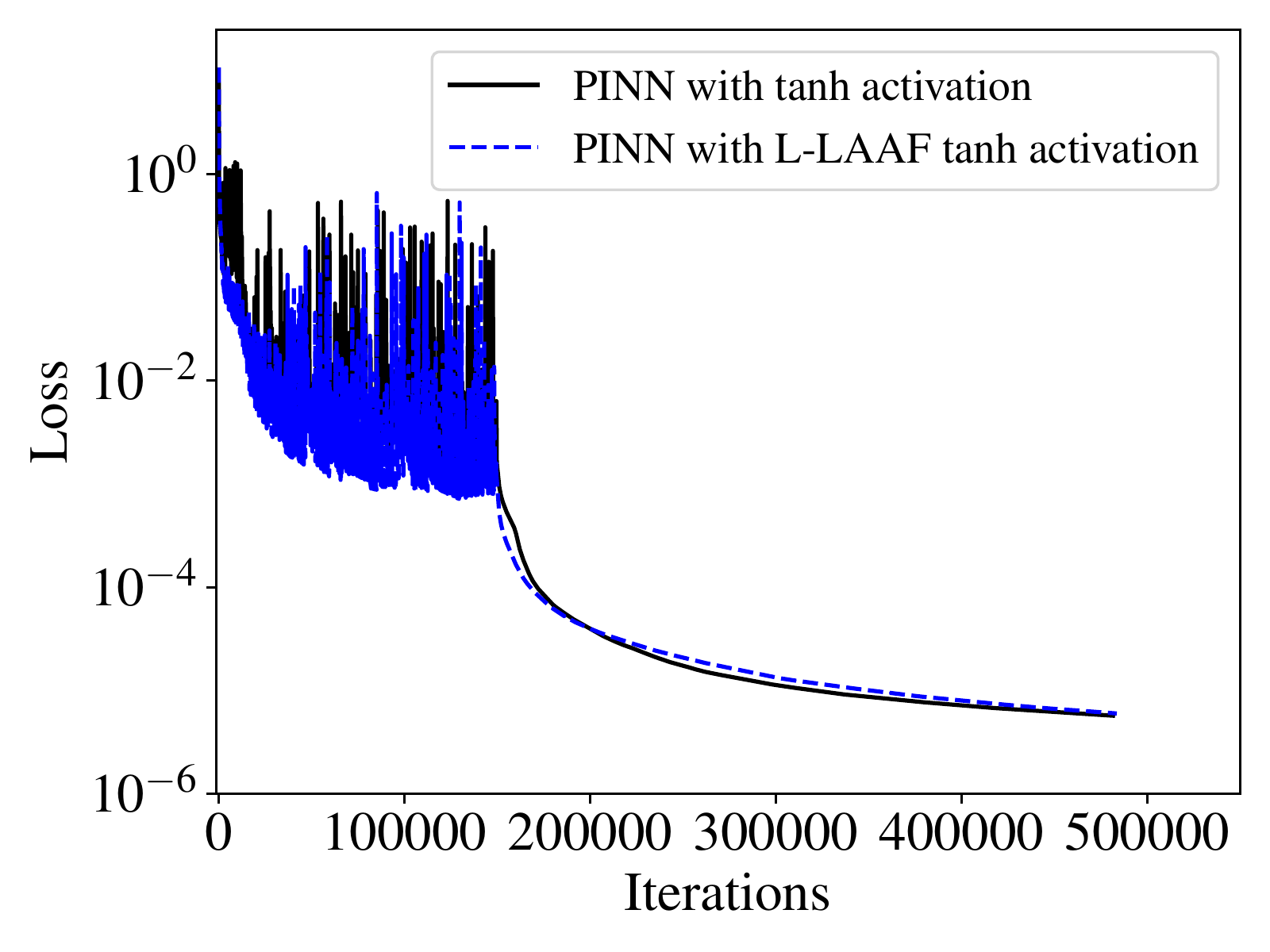} & & \includegraphics[height=170pt,trim={0.4cm 0.5cm 0.3cm 0.3cm},clip]{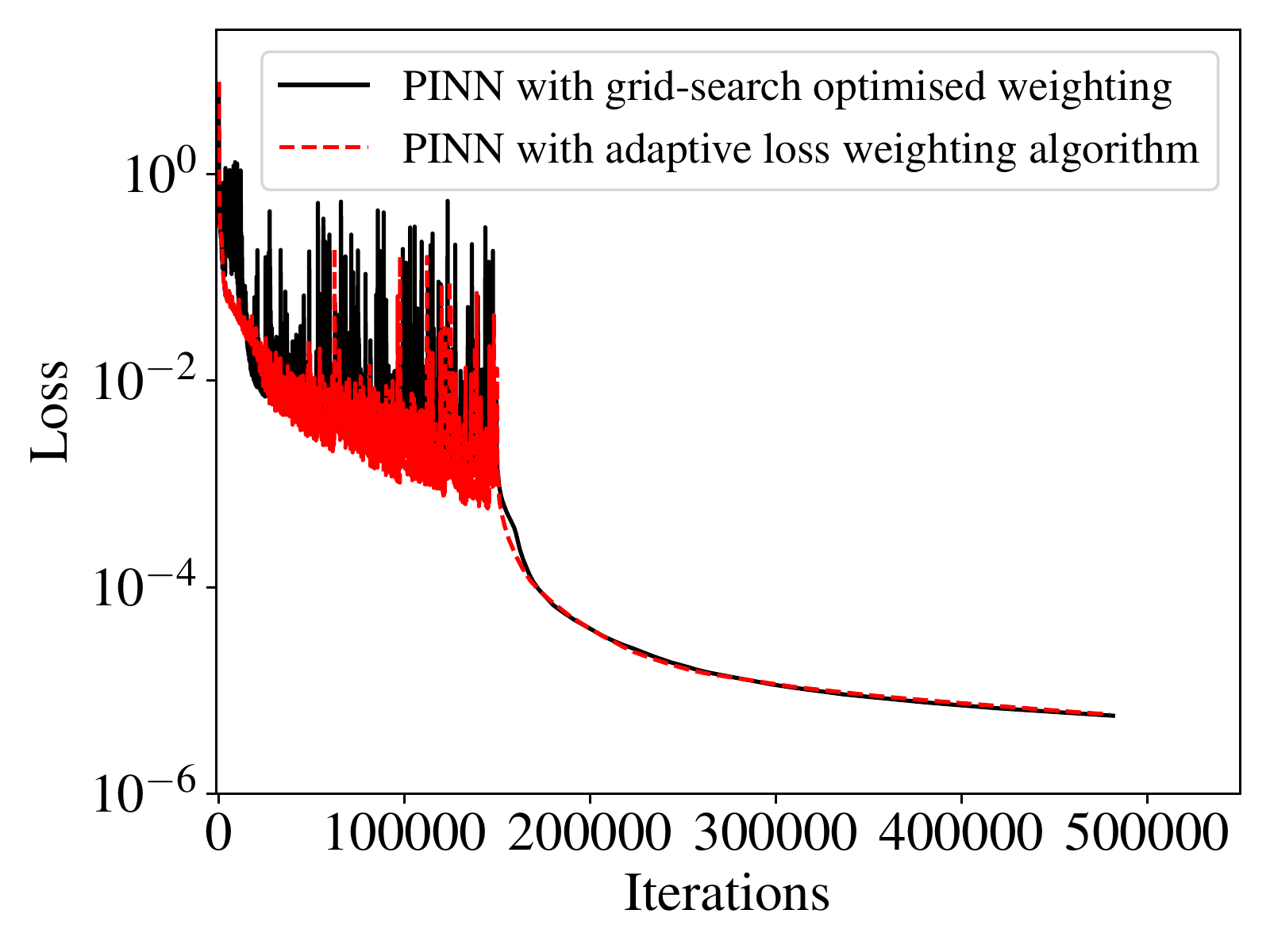} \\
  \end{tabular}
  \caption{Comparing the effect of convergence for PINN-DA-SA by applying (a) layerwise locally adaptive activation functions (L-LAAF) (b) an adaptive loss weighting algorithm.}
  \label{fig:adapt}
\end{figure}
Whilst a grid search approach was used to tune hyperparameters in this work, there have been many papers investigating the use of more robust methods. In this section, two approaches are investigated: adaptive activation functions and adaptive weights.

Adaptive activation functions, such as the layerwise locally adaptive activation function (L-LAAF) proposed in \citep{Jagtap:2020:2}, have shown promise at both accelerating convergence of PINNs, as well as improving model accuracy. For every layer, L-LAAF introduces a learnable tuning parameter, $a$ into the activation function to scale its gradient:
\begin{equation}
    \sigma^{*}(x) = \sigma (ax),
\end{equation}
where $\sigma^{*}, \sigma$ are the adaptive and base activation functions, respectively and $x$ is some input into the activation function. The effect of applying L-LAAF (with $\tanh$) to the periodic hill problem is presented in Table \ref{tab:hyp:act} and Figure \ref{fig:adapt}(a). For this problem, there was a minimal effect on the converged loss value. There does appear to be some improvement in the `spikiness' seen in the initial ADAM phase. However, during the second, long L-BFGS-B phase, this is no longer seen and the convergence speed is almost identical.

Adaptive weighting algorithms have been an increasingly promising area of research to make robust and reduce the difficulty of model weight tuning process. This is a particular challenge for PINNs, where there are often many loss terms in the equations. Figure \ref{fig:adapt}(b) demonstrates the advantage of using the adaptive loss weighting algorithm, as implemented in \citep{Sliwinski:2022}. Whilst the difference in the final loss between the model with optimised weights (from Table \ref{tab:weight}) and the model with adaptive weighting is minimal, the latter of these avoids all the intermediate grid-search steps used to find and select the optimal weights. This massively reduces the computational cost of finding the correct weighting and is thus the recommended approach moving forward.

\section{\label{sec:app:rste}Reconstructing pressure with PINN-DA-Baseline}
\begin{figure}[t]
    \centering
    \begin{tabular}{c}
    \begin{tabular}[b]{lc}
        \small (a) & \\
        & \includegraphics[width=0.85\textwidth,trim={0.5cm 0.6cm 0.25cm 0.4cm},clip]{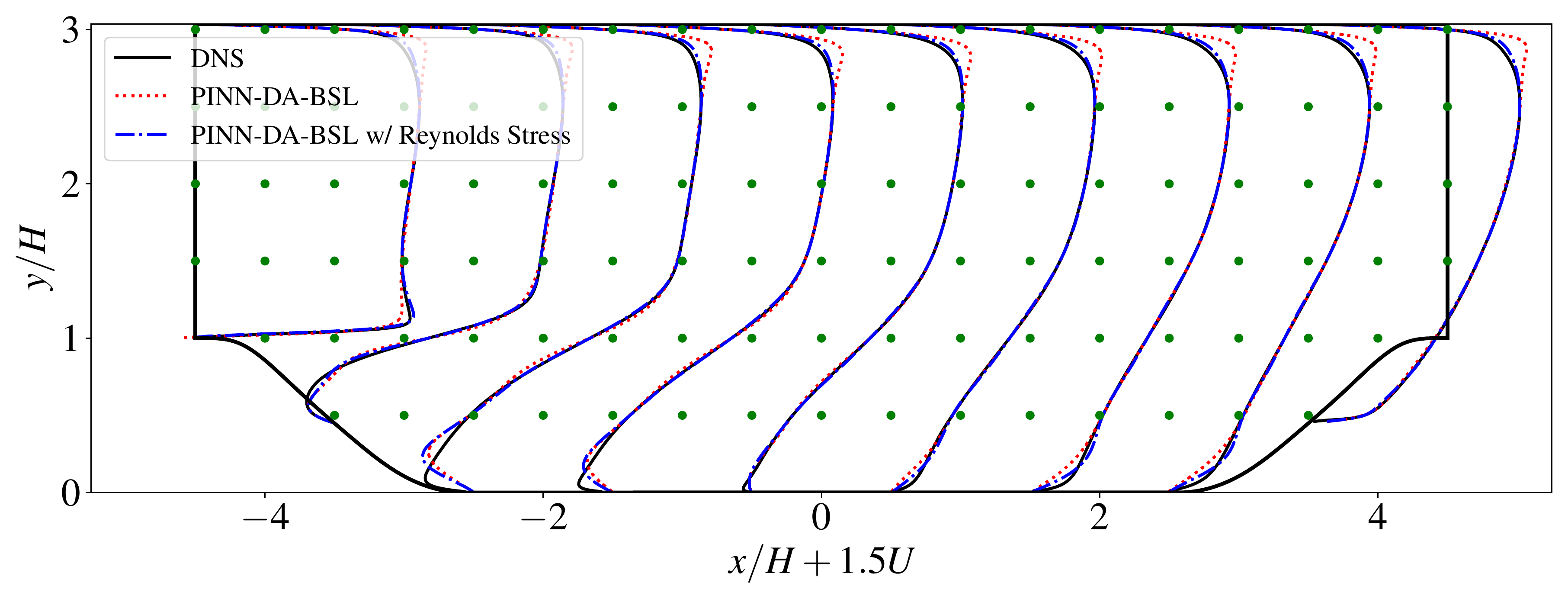}
    \end{tabular}\\
    \begin{tabular}[b]{lclclc}
        \small (b) & \hspace*{0.5cm}$\hat{P}$ & \small (c) & $P_{truth}$\hspace*{1.0cm} & \small (d) & $P_{err}$\hspace*{1.0cm}\\
        & \includegraphics[height=103pt,trim={0.5cm 0.6cm 4.3cm 0.35cm},clip] {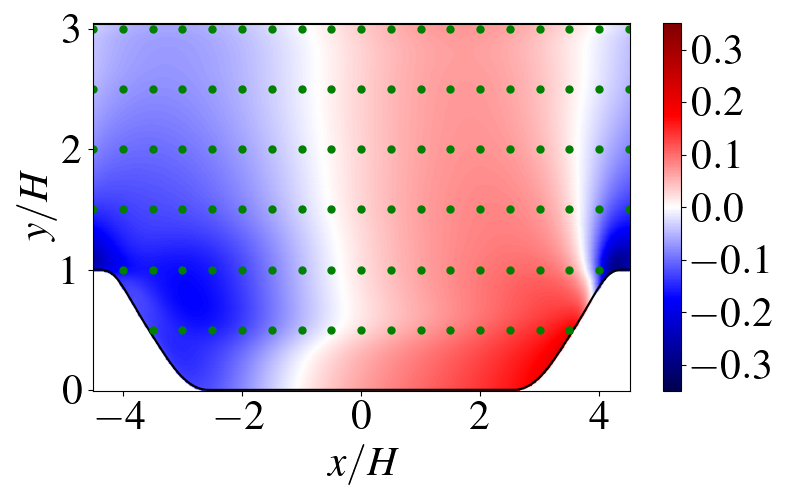} & & \includegraphics[height=103pt,trim={2.3cm 0.6cm 0.7cm 0.35cm},clip]{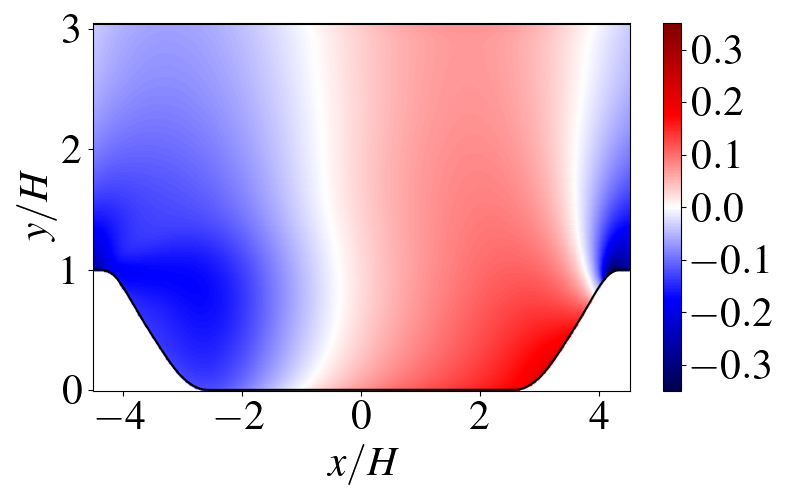} & & \includegraphics[height=104pt,trim={2.3cm 0.6cm 0.5cm 0.35cm},clip]{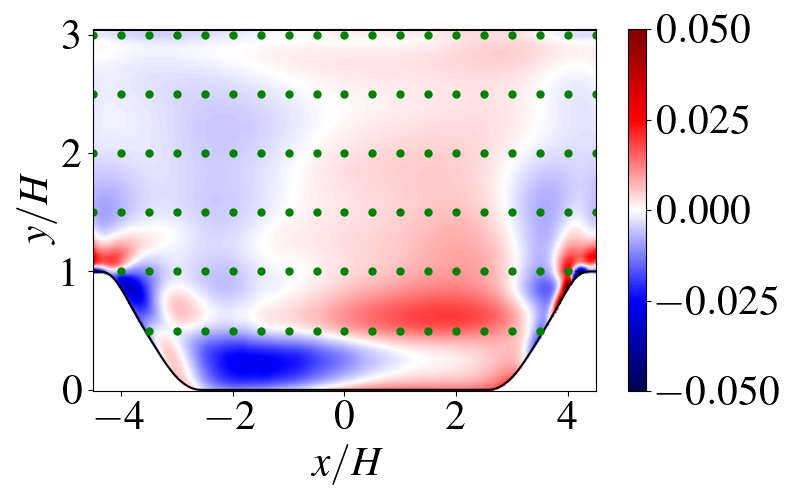}
    \end{tabular}
    \end{tabular}
    \caption{PINN-DA-Baseline mean flow reconstruction. (a) Comparison between the DNS, PINN-DA-Baseline and PINN-DA-Baseline (with Reynolds stress data) velocity profiles for the $U$ component. (b) Pressure field inferred from the PINN-DA-Baseline approach from sparse mean velocity and Reynolds stress data. (c) Absolute mean pressure error from PINN-DA-Baseline pressure reconstruction.}
    \label{fig:PH:PINNB:pres}
\end{figure}
Instead of applying the Helmholtz decomposition \eqref{eq:RANS:helm} with the PINN-DA-Baseline, as in Section \ref{sec:res:PINNB}, one can alternatively enforce the base RANS equations as presented in \eqref{eq:RANS}, using both high-fidelity first order statistics (mean flow) and second order statistics (Reynolds stress) ($U_{i}, \overline{u_i'u_j'}$) at discrete locations to reconstruct the flow field. This approach enables reconstruction of both mean velocity and Reynolds stresses fields, but also allows the inference of the pressure field. 

\begin{table}[b]
\caption{Weights used in PINN optimisation when both mean velocity and Reynolds stress data are provided}
\label{tab:weight:2}
\begin{ruledtabular}
\begin{tabular}{lcccccc}
\textrm{Weight}&
\textrm{$\lambda^{P}$}&
$\lambda^{D}(U_{i})$&$\lambda^{D}(\overline{u_i'u_j'})$&
$\lambda^{B}(U_{i})$ - Wall&$\lambda^{B}(\overline{u_i'u_j'})$ - Wall&
\textrm{$\lambda^{B}$ - Periodic}\\
\colrule
Value &
1 &
10 & 100 &
2.5 & 10 &
1\\
\end{tabular}
\end{ruledtabular}
\end{table}
%
Using the data spacing $\Delta L=0.5$, shows that this alternate PINN-Baseline approach produces comparable mean velocity reconstruction as the PINN-DA-Baseline, as seen in Figure \ref{fig:PH:PINNB:pres}(a). The $L_2$ error is $3.37\times 10^{-2}$ for this second approach compared with $3.60\times 10^{-2}$ for the former approach. Likewise the key trends discussed in Section \ref{sec:res:PINNB} all still hold. However, this alternate formulation has more output variables and fewer equations when compared to the approach used in the main paper.

Although both approaches achieve similar reconstruction error, this second approach requires second order statistics (Reynolds stress measurements) along with first order ones (mean flow field). However, this formulation can be used to reconstruct the pressure field, as shown before for laminar cases~\citep{Sliwinski:2022}. This is demonstrated in the turbulent case, as seen in Figure \ref{fig:PH:PINNB:pres}(b,c). This has a high potential for practical applications in order to obtain the pressure field without performing intrusive measurements. Examination of the RANS equations \eqref{eq:RANS} shows how provision of mean velocity and Reynolds stress data at discrete points defines all terms in the equations except for the pressure term. At these measurement locations, the equations are closed and the resultant neural network can infer a pressure field, both at data points but also away from them as well. This resultant pressure field has highest reconstruction error in the regions with highest pressure gradients in the flow, matching the trend seen with mean velocity error. The pressure reconstruction error is less correlated with the location of data points, than the mean velocity. It should be highlighted that the neural network pressure output must also include an integration constant (as the RANS equations only contain derivatives of pressure). To determine this constant and recover the actual, unshifted pressure field, one needs to fix the pressure at one point. This can be done during training (by providing pressure data at a single point) or as in this case, the pressure field is shifted in post processing. Taking the pressure data at a singular point, one can find a constant, such that by adding it to the model pressure at that point, the predicted pressure exactly matches the measurement. This constant can then be applied to the full solution.
%

\section{Laminar Cylinder Flow\label{sec:app:laminar}}
\begin{figure}[t]
  \centering
  \begin{tabular}[b]{clc}
    & & \hspace*{0.3cm}DNS \\
    & \small (a) & \\
    & & \parbox[c]{211pt}{\includegraphics[height=130pt,trim={0.4cm 1.9cm 3.8cm 0.2cm},clip]{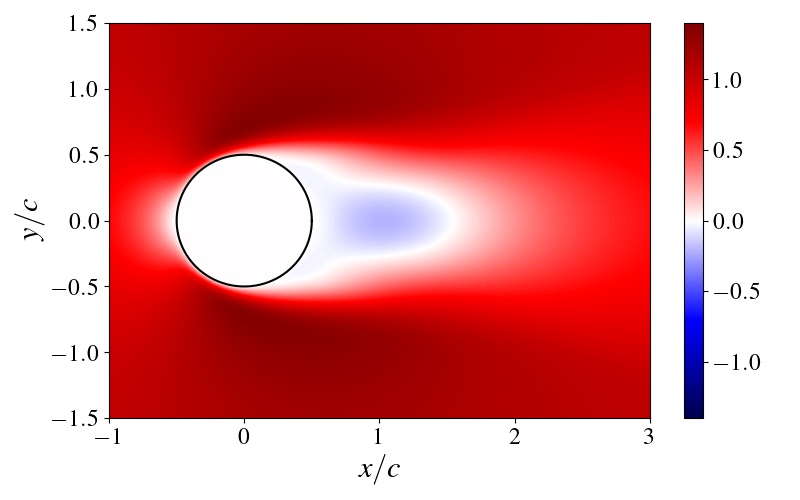}}
  \end{tabular}
  \\ \vspace{0.5cm}
  \begin{tabular}[b]{clclc}
    & & \hspace*{0.2cm}Variational-DA & & PINN-DA-Baseline\hspace*{1.4cm} \\
    & \small (b) & & \small (c) & \\
    $\hat{U}$ & & \parbox[c]{211pt}{\includegraphics[height=130pt,trim={0.4cm 1.7cm 3.8cm 0.2cm},clip]{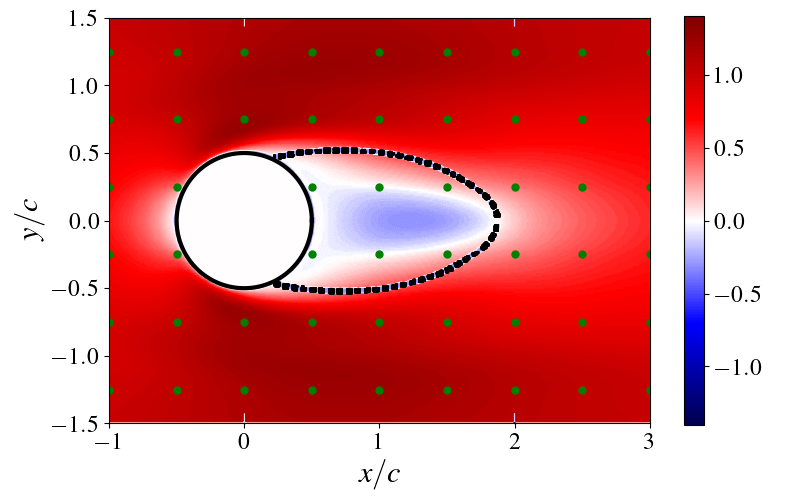}} & & \parbox[c]{211pt}{\includegraphics[height=130pt,trim={2.6cm 1.85cm 0.7cm 0.35cm},clip]{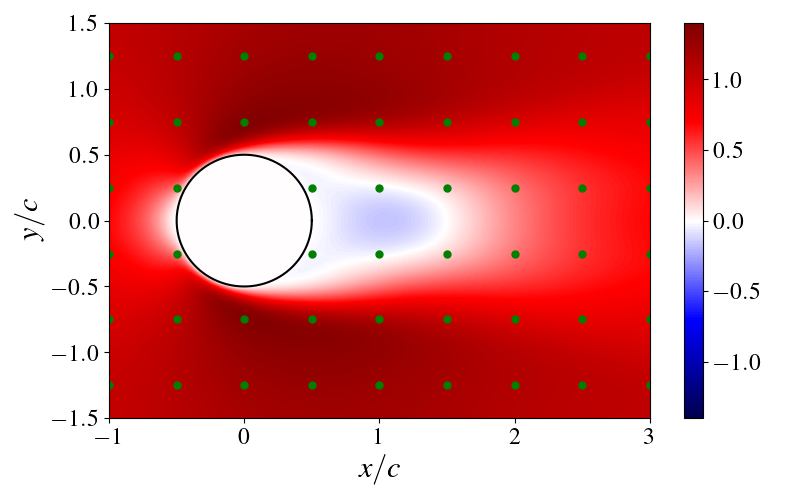}} \\
    & \small (d) & & \small (e) & \\
    $U_{err}$ & & \parbox[c]{211pt}{\includegraphics[height=147pt,trim={0.4cm 0.2cm 3.8cm 0.35cm},clip]{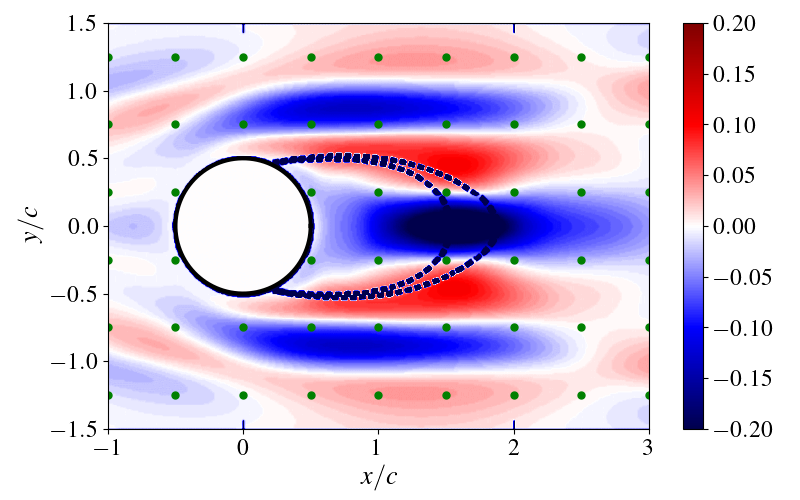}} & & \parbox[c]{211pt}{\includegraphics[height=147pt,trim={2.6cm 0.45cm 0.7cm 0.35cm},clip]{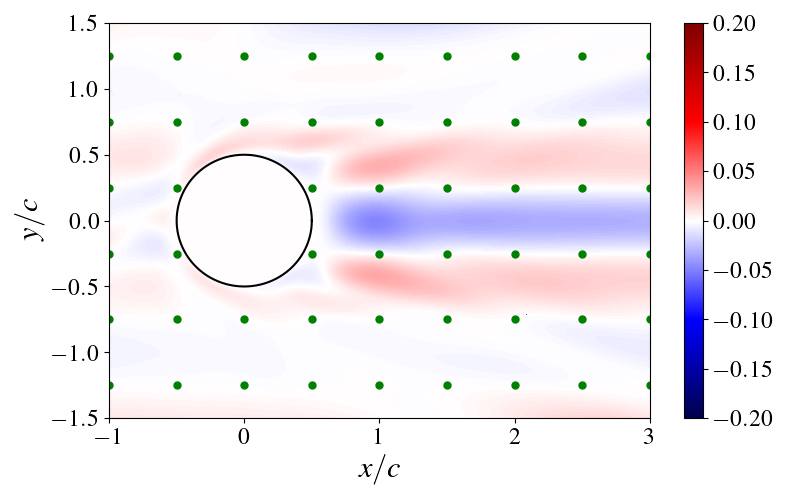}}
  \end{tabular}
  \caption{Comparison of the $U$ (x-component) mean velocity component reconstructions for $d=0.5$ data resolution with DNS (top). The variational-DA approach (left) and the PINN-DA-Baseline (right) figures are taken from \citep{Foures:2014} and \citep{Sliwinski:2022}, respectively. (a) - $U$ truth. (b,c) - $U$ prediction. (d,e) - Absolute $U$ error.}
  \label{fig:lam:comp}
\end{figure}

Both the variational-DA~\citep{Foures:2014} and PINN-DA-Baseline~\citep{Sliwinski:2022} techniques (without the use of turbulence model augmentation) have already been successfully demonstrated on a laminar circular cylinder case. However, a quantitative comparison has not been completed, which is presented in this section.

A direct numerical simulation of a 2D circular cylinder flow at $Re=150$ is performed in order to generate the high fidelity database, which is used to extract sparse mean velocity measurements for use in the above data-driven techniques. The details of the simulation setup and methodologies can be found in ~\citep{Foures:2014} and ~\citep{Sliwinski:2022} for the variational and PINN case respectively. Figure \ref{fig:lam:comp}(a) shows the time-averaged mean $U$ field for the circular cylinder. For this work, the data spacing, $d = 0.5$, on a rectangular grid is used to compare approaches as a comparison case. The aforementioned papers also contain results of differing data resolutions.

This direct comparison of both variational-DA and PINN-DA approaches was used as a benchmark for what the turbulent applications can achieve (as shown in the main paper). These results will demonstrate that for laminar flow, the PINN-DA approach flow reconstruction can match and improve the variational-DA reconstructions. 

Figure \ref{fig:lam:comp} shows the reconstructed $U$ field (b,c) and the prediction error (d,e). This highlights several trends, which are consistent across all velocity components and were also seen in the turbulent cases. Firstly, the reconstructed mean velocity fields are very similar to the true fields and accurately capture the expected physical features, such as the symmetrical wake and the stagnation at the front of the cylinder. Additionally, a key pattern emerged from this result that was consistent across both the laminar and turbulent cases. Both the variational-DA and PINN-DA approaches have almost identical reconstruction error fields. These occur between datapoints, in the wake and recirculation regions and near the wall. Finally, as one moves away from the cylinder (in $y$), the error reduces as the flow becomes more uniform.

Whilst the reconstruction error fields are comparable, it is clear that the absolute error for the PINN-DA-Baseline is much smaller in magnitude. The variational-DA has a maximum error magnitude of $0.3$ (for absolute $U$ error), whereas for PINN-DA it is around a third of this at $0.11$. This comparison is important as it suggests, all else equal for a laminar cylinder flow, the PINN-DA produces better results. This is most apparent when comparing the length of the wake. Whilst the variational approach predicts the wake recirculation region closes at $x=1.85$, the PINN approach closes at $x=1.55$ (compared with $x=1.55$ for the DNS flow field). Furthermore, the decay in error, observed as one moves away from the body, is much more apparent in the PINN-DA results. This may indicate the fundamental nature of PINNs and their construction, allows them to better generalise these simple uniform regions.
\clearpage
\bibliography{bibliography}


\end{document}